# Automated Gaze-based Behavioral Segmentation and Temporal Representation for Bridge Inspection in Unconstrained 3D Environments

Daniel Jimenez Gil [a,*], Haosen Zhang [b], Zixin Wang [a], Mohamad Alipour [a]

a. Department of Civil Engineering, The Grainger College of Engineering, University of Illinois Urbana-Champaign, Urbana, IL 61821, USA
b. Department of Economics, College of Liberal Arts & Sciences, University of Illinois Urbana-Champaign, Urbana, IL 61821, USA

* Corresponding Author

## Abstract

Visual bridge inspection is a knowledge-intensive task in which inspectors coordinate visual search, spatial navigation, structural reasoning, and defect identification and documentation. It is a central maintenance task for bridges and a key basis for safety assessments, yet its results are susceptible to individual subjectivity. While eye-tracking-based behavioral studies quantify underlying processes, existing research often imposes restrictive simplifications to reduce environmental complexity, thereby compromising ecological validity. This study proposes an automated data analytics framework for converting multimodal inspection data into an inspection mode time series. Unconstrained 3D gaze, head-movement, drone navigation, and scene geometry data are segmented into temporal windows and classified into three functional modes: global scanning, local inspection, and navigation. The resulting temporal representation enables the extraction of interpretable behavioral descriptors, including transition probabilities, dwell times, transition entropy, fixation measures, and spatial revisit metrics. A feasibility study using a virtual bridge inspection platform demonstrates that the proposed representation captures meaningful differences in inspection strategy and reveals exploratory relationships with inspection performance. This study contributes a framework for human-informed computer-aided infrastructure inspection systems, inspector training, and data-driven assessment of constructed facilities.

# 1 Introduction

## 1.1 Motivation

To address the monumental task of maintaining transportation infrastructure, agencies rely heavily on the preventative and predictive capabilities of periodic bridge inspections. While visual inspection remains the primary method for assessing safety and structural integrity, its efficacy is inherently limited by individual subjectivity in inspector experience and judgment, significant inter-inspector variability, and a lack of standardized reproducibility. This remains true despite the extensive training and licensing procedures required for certified inspectors. For example, a study demonstrated that 95% of primary element condition ratings for individual bridge components vary within two rating points of the average, while only 68% vary within a single point [1]. Such reliance on human judgment introduces uncertainties that can lead to inconsistent condition ratings, which in turn may result in misallocated maintenance resources.

To mitigate these limitations and better understand the underlying cognitive processes of inspectors, recent research has increasingly utilized eye-tracking and behavioral analysis [2,3]. By capturing gaze patterns, researchers can quantify the visual search strategies and decision making heuristics that differentiate inspection strategies and identify approaches associated with successful inspection. However, the application of eye-tracking in complex, unconstrained 3D environments is often hindered by the limitations of current analytical frameworks. Most existing studies rely on restrictive experimental controls, such as 2D image-based simplifications or labor-intensive manual area-of-interest (AOI) definitions, which diminish the ecological validity of the findings. Manual AOI mapping, in particular, not only introduces potential researcher bias but also fails to scale to large-scale infrastructure assets with complex geometries. Consequently, there remains a technical gap in automatically transforming unconstrained 3D gaze data into semantically meaningful and functionally structured information that can be objectively linked to performance outcomes.

Furthermore, as the industry adopts digital platforms, such as desktop-based environments and Virtual Reality (VR) visualizations for inspector training and remote assessment, the ability to automatically quantify user attention within these models is essential. There is a critical need for automated and scalable methods that can segment and characterize complex 3D inspection behaviors. Developing a pipeline that transforms raw gaze data into high-level behavioral descriptors is a key step toward creating objective benchmarks for inspection quality and informing the next generation of automated inspection support systems. This need is directly connected to the broader automation of inspection and maintenance workflows for constructed facilities. As bridge owners and engineering organizations increasingly adopt virtual inspection environments, UAS-assisted data collection, and digital models for infrastructure management, there is a growing need for methods that can automatically transform human inspection behavior into measurable and reusable process data. In this context, gaze and movement data are not treated only as human-factors measurements, but as information sources for computer-aided inspection assessment, training, and future automated inspection systems.

### 1.2 Bridge inspection as a visual search task

A review of the literature shows the scarcity of studies examining the cognitive processes and visual search framing specific to bridge inspections. Nevertheless, principles from the broader visual search literature, construction hazard research, and aviation studies can be adapted for the theoretical foundation of this study. In particular, the process of finding targets in a complex scene through distinct attention mechanisms is modeled by Wolfe's Guided Search framework [4]. This framework can provide a basis for understanding how large environments are scanned by bridge inspectors amid high levels of noise and distractors while searching for defects such as cracks, corrosion, or delamination. Based on this model, experienced inspectors build a global priority map to help guide a focused search for features of interest. They leverage both bottom-up guidance (the physical salience or visual contrast of features) and top-down guidance (goals, experience, and knowledge of typical failure modes) in this process. In bridge and aviation inspections, this theory is echoed by findings that experts tend to have saliency maps reflecting top-down reasoning (i.e., more organized and directional), whereas non-experts may rely more heavily on bottom-up features [3,5,6].

Beyond visual search, gaze is guided significantly by the structure of the inspection task itself [7–10]. The inspector's goal is not simply to find a defect on a bridge, but also to identify it and decide whether it is structural or superficial. Therefore, the inspection process comprises multiple subtasks that are required to gather information, process it, and make decisions. Information processing is explained in three stages by Endsley's Situational Awareness (SA) model, namely perception, comprehension, and projection [11]. When applied to bridge inspections, broad scanning of the structural environment (global SA) is necessary for the initial perception of potential anomalies. Subsequently, a focused visual attention on specific elements (local SA) is needed for the comprehension of a defect's severity and the projection of its structural impact. Luo et al. [12] proposed a framework for measuring the magnitude and levels of SA in 3D environments based on physiological markers, such as pupil dilation. However, it is difficult to isolate the measurement of SA when applying these methods to an unconstrained 3D environment.

Finally, a major difference when switching from a constrained environment, such as a 2D image-based task, to an unconstrained 3D environment is the additional cognitive load imposed by navigation. Enders et al. [13] conducted a divided-attention experiment showing that navigation and gaze slow down during periods of cognitive processing, because navigation subtasks (such as collision avoidance, path planning, and coordination) compete for the user's limited cognitive capacities. Divided attention between the operational task of navigation (piloting a drone) and the cognitive task of structural assessment is similarly required in bridge inspections. Overall, an inspector’s cognitive processes in a 3D environment can be described as a dynamic sequence of global scanning and local inspection activities facilitated by navigation throughout the 3D environment and informed by the inspector’s goals, knowledge, experiences as well as the visual characteristics of the inspection target.

### 1.3 Eye tracking for engineering cognitive processing

Eye tracking has increasingly been used to study cognitive processing in engineering tasks because gaze provides a measurable proxy for how users allocate attention, search for information, and respond to task demands. Prior studies in construction safety, aviation, quality control, and visual inspection have linked fixation-based features to hazard recognition, situational awareness, expertise, workload, and inspection performance [12,14–19]. In construction safety, Hasanzadeh et al. [14,20] studied factors affecting the detection of safety hazards during construction safety inspections, including work experience and previous exposure to hazards, and linked these markers to gaze features, such as dwell time, fixation count, time to first fixation, and number of returns. Their qualitative analysis of the gaze heatmap and fixation path revealed qualitative differences between trained and untrained inspectors. Similarly, Jeelani et al. [21] examined gaze-based indicators of hazard recognition performance, while Luo et al. [12] proposed measures of situational awareness in construction workers. Together, these studies show that gaze can provide useful indicators of engineering cognition, particularly when gaze features are linked to task-related cognitive constructs rather than treated as isolated visual metrics.

Inspection studies have also shown that gaze behavior reflects differences in inspection strategy and expertise [3,5,22,23]. Aust et al. [5,22] studied aircraft engine blade inspection and found that inexperienced inspectors produced more dispersed heatmaps and were more frequently drawn toward visually salient distractors, suggesting greater reliance on bottom-up attention. Experienced inspectors, by contrast, tended to perform a broader, more holistic inspection before engaging in a detailed search. Liu et al. [3] reported similar expertise-related differences in bridge analysis using eye-tracking data from inspectors examining a finite element bridge model. This study allowed inspectors to freely rotate the model in an unconstrained environment, while the fixation analysis was primarily qualitative, relying on observations from gaze heatmaps, reflecting the difficulty of applying user-centric analysis to unconstrained 3D gaze data. In follow-up research, Liu et al. [24] used interaction data, such as navigation and information-gathering actions, with machine learning to cluster inspection behavior and estimate inspection reliability. While fixation dynamics were not used as a primary basis for behavior classification, this work represented a step toward automated behavioral characterization.

Because inspection unfolds over time, temporal gaze modeling is especially relevant. Hayashi [6] utilized hidden Markov models (HMMs) to identify recurring attention patterns amongst pilots and showed that gaze states varied across phases of the landing approach and between experienced and inexperienced pilots. Ulutas et al. [25] applied a similar HMM approach to quality control inspections and reported that experienced inspectors first developed a broader view of the inspection scene before shifting to a detailed search. These studies demonstrate that gaze sequences can reveal latent behavioral states that are not captured by aggregate fixation statistics alone.

Transferring these gaze-modeling approaches to bridge inspection is challenging because the assumptions that make gaze analysis tractable in many prior studies do not hold in 3D inspection environments. In aviation, quality control, and image-based inspection studies, an object-centric approach, where the targets are known in advance, works with manual AOIs. Bridge inspections, by contrast, require unconstrained movement of inspectors through a large 3D environment, continuously changing viewpoints, and attending to many structural elements that may contain defects. As a result, manually defined AOIs become difficult to scale, and aggregate gaze measures provide limited insight into how inspectors coordinate visual search, navigation, and assessment over time. Li et al [26] addressed part of this challenge by creating a virtual platform for training and assessing bridge inspectors that allowed for unconstrained movement and realistic visual exploration, focusing mainly on analyzing the inspector's navigation abilities, with limited exploration of visual attention. However, the defects presented to bridge inspectors were limited to cracking and were randomly distributed throughout the bridge, not following any natural pattern. Therefore, experienced inspectors may not necessarily have an advantage by knowing common failure modes. This platform was later used by Dalilian et al. to evaluate the effects of AI-assisted bridge inspection on human detection error, and found a significant association between drone piloting experience and inspection accuracy [2]. Nevertheless, these studies did not consider the temporal dynamics of gaze behavior, which are important markers of task for highly dynamic environments.

### 1.4 Objectives and contributions

The primary objective of this study is to develop and evaluate a user-centric, automated behavioral segmentation framework to characterize inspection behavior in unconstrained 3D environments. The broader goal is to support automated analysis of bridge inspection workflows during infrastructure operation and maintenance by converting complex 3D inspection activity into structured, quantitative, and interpretable process descriptors. The proposed framework represents inspection as a sequence of three functional behavior states: global scanning, local inspection, and navigation. Global scanning corresponds to broad visual exploration of the bridge environment, during which inspectors develop an overall understanding of the scene and identify salient regions for further attention. Local inspection occurs when inspectors focus on a specific structural element to determine whether an observed feature constitutes a defect. Navigation captures periods in which cognitive effort is primarily governed by movement through the environment and control of the drone, rather than by direct structural assessment. Together, these three states provide a functional representation of how inspectors alternate among scene exploration, detailed evaluation, and movement control during an inspection task.

To classify these behavioral states, the proposed approach applies temporal windowing to gaze data, extracts features from each window, and uses them in a supervised learning setting to classify the inspector's behavior as global scanning, local inspection, or navigation. The resulting sequence of classified windows forms an inspection mode time series that captures the temporal evolution of inspection behavior. This representation enables signal processing techniques to be applied to behavioral data, allowing higher-level descriptors of inspection strategy and performance to be extracted.

This approach offers several advantages over traditional gaze analysis methods. Conventional methods often require predefined areas of interest (AOIs), constrained inspection paths, or simple aggregate gaze measures, such as fixation count and gaze angular velocity. While such measures are useful, they offer limited insight into how inspection behavior evolves over time throughout a task, particularly in unconstrained 3D environments. In contrast, the proposed method does not require manually defined AOIs and can incorporate multiple gaze, head movement, and navigation features within a unified temporal framework.

To demonstrate the application of the proposed framework, VISTA, a virtual bridge inspection platform that integrates bridge inspection simulation with eye movement data analysis, is introduced and made available under an open-source license on GitHub [27]. A small-scale feasibility study was conducted in the VISTA platform to examine whether descriptors derived from the inspection mode time series are associated with inspection performance. This study evaluates 16 behavioral descriptors and 9 performance metrics through descriptor-metric correlation analysis. In addition, inspection redundancy is quantified using an area-revisit heatmap, introduced in this study as a descriptor of repeated spatial coverage during bridge inspections. Overall, the main contributions of this study are as follows:

1. A virtual bridge inspection platform, VISTA, that integrates 3D inspection simulation with synchronized eye-movement data collection and analysis.
2. A tri-modal behavioral framework that represents bridge inspection as transitions among global scanning, local inspection, and navigation, enabling signal-processing-based extraction of behavioral descriptors from classified inspection states.
3. An automated temporal-windowing pipeline including a machine learning model for classifying inspection behavior in unconstrained 3D environments without relying on manually defined AOIs.

To the best of the authors' knowledge, this is the first study to represent bridge inspector behavior in terms of functional modes by classifying gaze and movement data and to extract performance-related features from the resulting inspection mode time series and correlating them with inspector performance.

# 2 Methodology

## 2.1 Overview

The study consists of four main components. First, the Visual Inspection Simulator for Tracking Attention (VISTA) was developed as a bridge inspection simulation platform with integrated eye-tracking and movement data collection and analysis. Second, participants inspected a virtual bridge in VISTA under a time constraint, searched for defects, and completed a skill evaluation survey. Third, inspection modes were labeled by a group of annotators, and machine learning classifiers were trained to classify inspection behavior from gaze and movement features. Finally, descriptors were extracted from the resulting inspection mode time series and evaluated for their association with inspection performance. Figure 1 provides an overview of the proposed methodology, including these four components. This workflow frames bridge inspection as a computer-aided inspection process in which simulation, synchronized sensing, machine-learning classification, and behavioral descriptor extraction are integrated to support objective assessment of inspection quality.

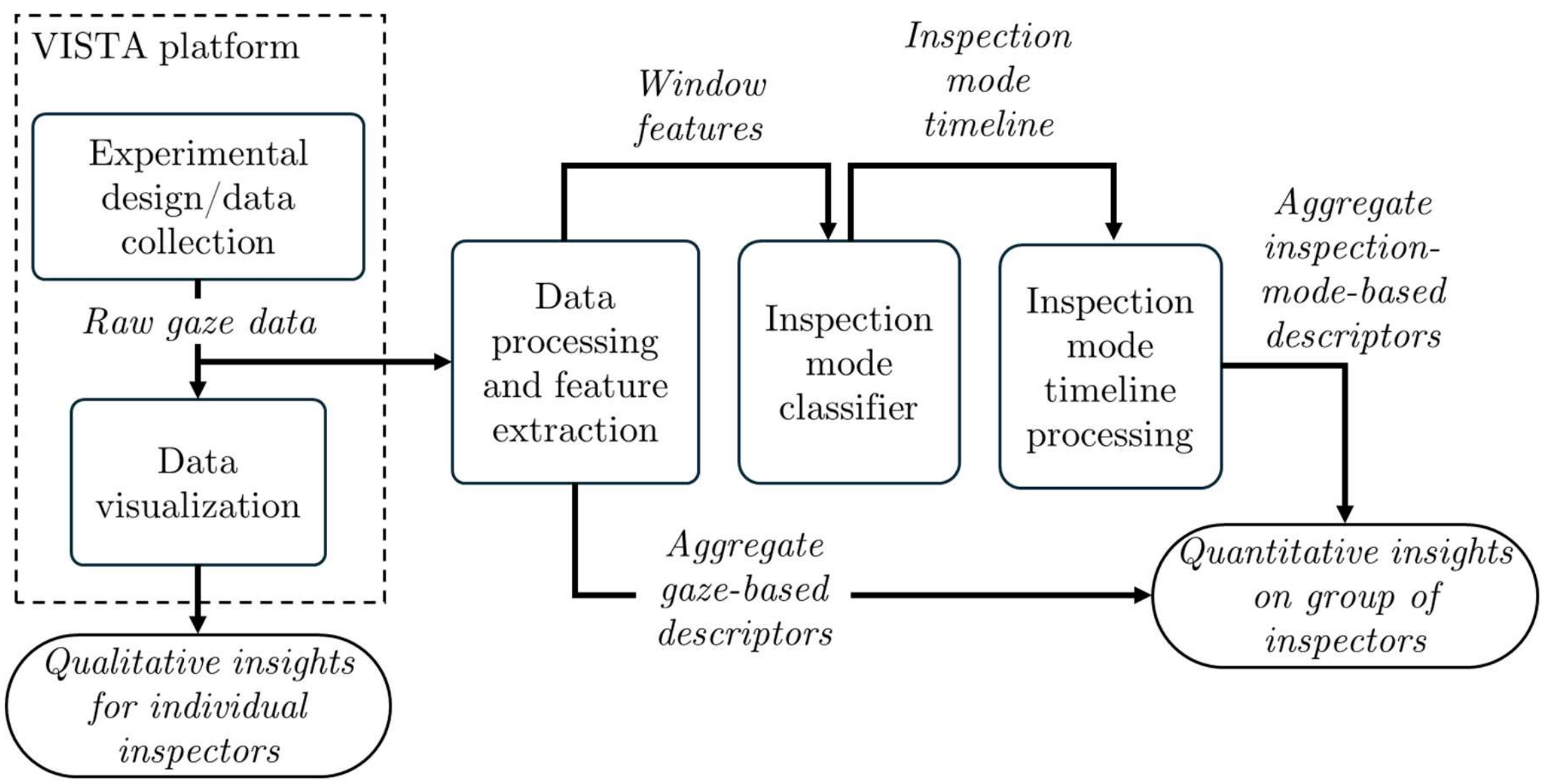

***Fig. 1.*** *Overview of the proposed methodology for qualitative and quantitative analysis of bridge inspection behavior.*

In the proposed methodology, the VISTA platform supports both experimental design and synchronized data collection, including raw eye-tracking and drone-movement data. The platform also generates visualizations that enable qualitative evaluation of individual participants' behavior. For quantitative analysis, the proposed pipeline applies temporal windowing to the raw data. Each window is converted into a set of gaze, head-movement, and navigation features informed by prior work on task prediction and cognitive-state analysis [8,14,15,19,28]. The features were used to classify the windows into one of three inspection modes: global scanning, local inspection, or navigation. The classified sequence of windows then forms an inspection mode time series, which represents how inspection behavior evolves over time. From this time series, aggregate descriptors are calculated to characterize the full inspection session. In this study, window-based features refer to measurements extracted from short temporal segments and used for inspection-mode classification, whereas aggregate descriptors refer to session-level measures derived from the complete inspection mode time series and used to evaluate inspection strategy and performance.

## 2.2 VISTA platform

### *2.2.1 Platform design*

The Visual Inspection Simulator for Tracking Attention (VISTA) platform was developed using the Unity 3D engine to support bridge inspection simulation, synchronized data collection, and visualization. The Python programming language was used for data processing and feature extraction. The Tobii Pro Unity Software Development Kit (SDK) [29] was used to interface with the Tobii Eye Tracker 5, a screen-based eye tracker, which was used for data collection. Figure 2 shows the inspection interface presented to participants during data collection. The platform supports both first-person and third-person views; however, this study used a full-screen first-person view from the drone camera to reduce visual distractors and approximate a remote bridge inspection scenario in which the inspector navigates a drone-mounted camera through the bridge environment.

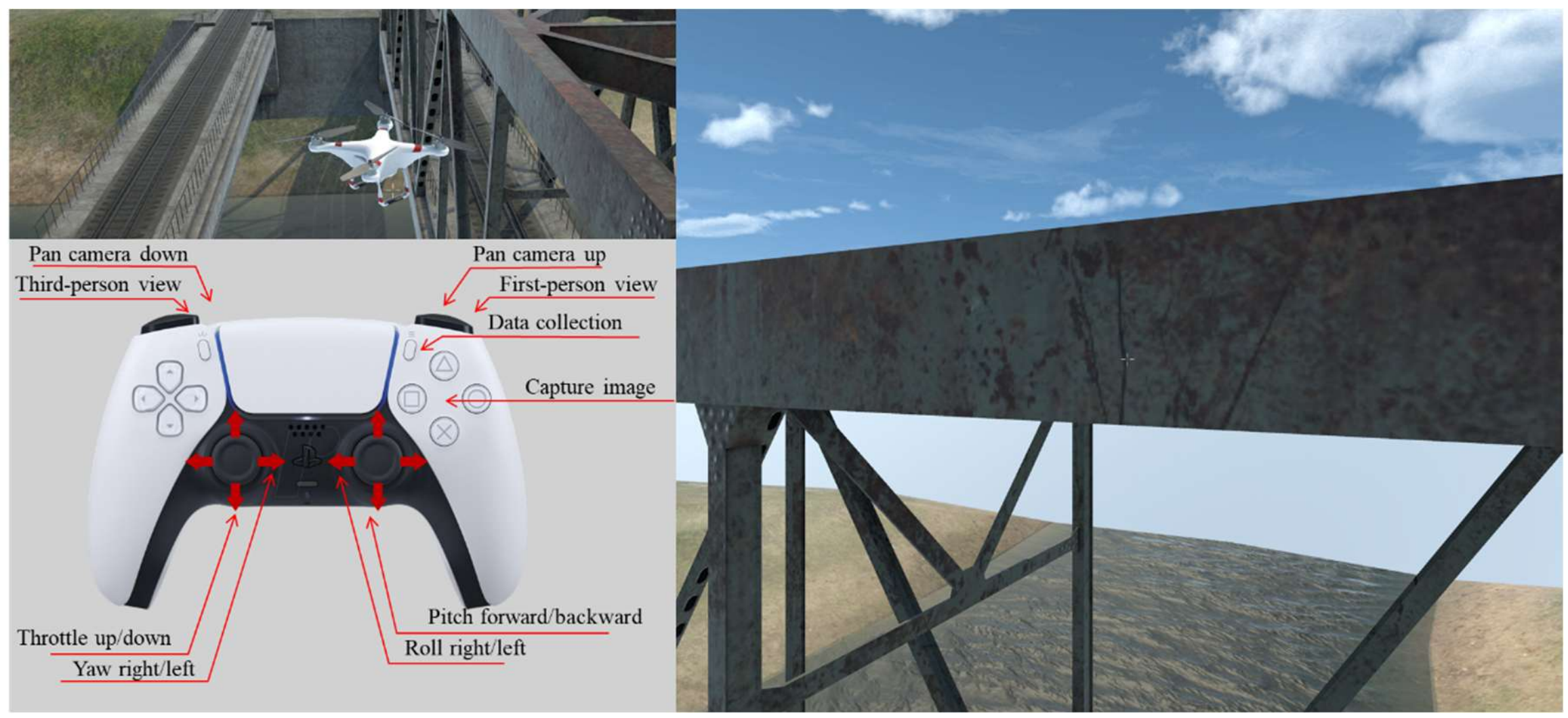

***Fig. 2.*** *View of the VISTA platform data collection module. Controller used for drone movement (Bottom-left). Drone in third-person view performing bridge inspection (Top-left) and the inspector's view in first-person (Right).*

At each simulation time step, VISTA records synchronized eye-tracking, camera, and movement data. Figure 3 illustrates the raw data collected at every time step in the simulation. This data is exported in Exportable Markup Language (XML) format at 60 frames per second. The camera position and orientation are recorded in Unity world coordinates and represents the location and viewing direction of the virtual drone camera in the simulation. The head position is estimated from the screen-based eye tracker and represents the user's viewing position relative to the display. The 2D eye gaze data is recorded relative to the upper left corner of the screen, while the 3D eye gaze points are computed by projecting a ray from the estimated head position towards the 2D eye gaze location. When the gaze ray does not intersect with the bridge model or the surrounding environment (e.g., when looking at the sky), the 3D eye gaze point is assigned a null value.

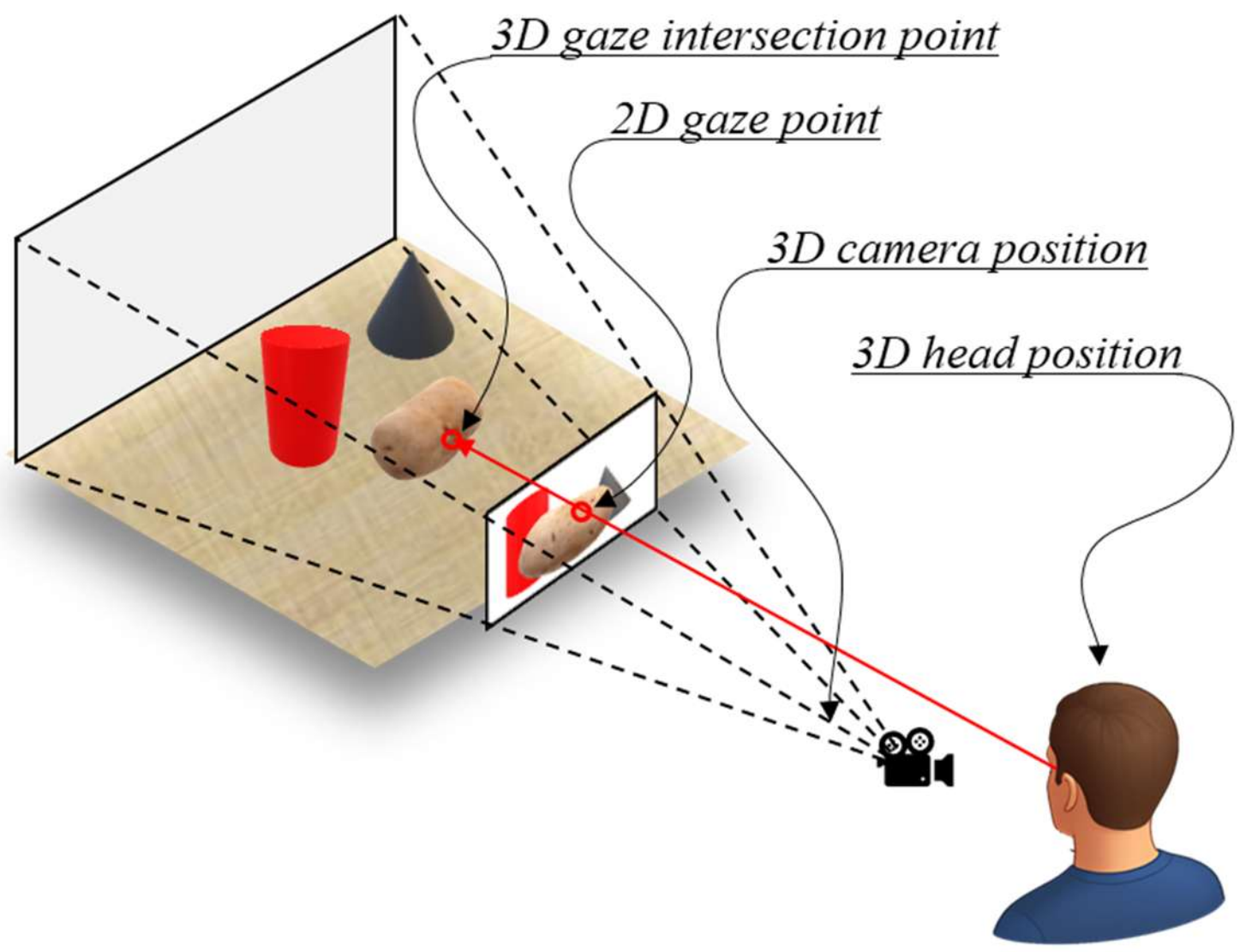

***Fig. 3.*** *Visualization of raw gaze, camera, and movement data extracted during the data collection.*

### *2.2.2 Data pre-processing*

Raw gaze data was pre-processed following the velocity-threshold identification fixation filter (I-VT) technique [30]. The preprocessing procedure includes low-pass noise filtering using a 100 ms window, blink detection, gap interpolation, and fixation classification. Gaps in gaze data longer than 75 ms were treated as blinks, while gaps shorter than 75 ms were filled using bilinear interpolation. Fixations were identified by implementing the I-VT fixation filter with a velocity threshold of 20 deg/s and a minimum fixation duration of 60 ms. Eye movement velocity was calculated using the method described by Duchowski [16]. For each valid gaze sample, the gaze vector was defined as

$$v_i = p_i - h_i \quad (1)$$

where $p_i$ is the 3D gaze intersection point and $h_i$ denote the estimated head postion at time step $i$ . Unlike the original formulation, which uses the average head position over a time period, this study used the instantaneous head position at each time step. This modification was necessary because the virtual camera and inferred viewpoint can change rapidly during drone navigation, resulting in erroneous angular velocity estimates when using average head positions.

The instantaneous visual angle between consecutive gaze vectors is calculated as

$$\theta_i = \cos^{-1} \frac{v_i \cdot v_{i+1}}{||v_i|| \, ||v_{i+1}||}. \tag{2}$$

A five-tap velocity filter was then applied following Duchowski [16], to estimate the angular velocity

$$\dot{\theta}_i = \frac{1}{\Delta t} \sum_{j=0}^{k} \theta_{i+j} H_j , \qquad i \in [0, n-k), \tag{2}$$

where $H_j$ refers to the unnormalized fixation filter represented by the discrete array {1, 2, 3, 2, 1}, with filter length k, $\theta_i$ is the instantaneous visual angle from Eq. (1), $\Delta t = t_{i+k} - t_i$ is the time difference between the sampling period $i$ and $k$, and $n$ is the total number of samples with a gaze intersection point.

Velocity-based filters were selected because they provide a simple, user-centric criterion for identifying fixations that does not require predefined areas of interest or environment dependent calibration [31]. This is important for unconstrained 3D bridge inspection, where object-centric approaches, such as area of interest filters (I-AOI) would necessitate extensive manual labeling. While user-centric but dispersion based filters such as the dispersion threshold fixation filter (I-DT) [31,32], would require a varying threshold depending on the environment. This setup may bias results toward larger or more densely labeled objects [31]. In contrast, velocity-based filters can be applied consistently across participants and inspection paths.

However, velocity-based fixation filtering has limitations in 3D environments. Because the classification is based primarily on angular change, gaze points that are close in visual angle but lie on different depth planes may be grouped into the same fixation. This limitation is less problematic in convex environments or short-duration viewing contexts but becomes more important over long inspection periods in complex 3D scenes. For this reason, the fixation filter is mostly suited to compute features within short temporal windows rather than to define long-duration semantic fixation patterns over the entire inspection. A more advanced treatment of fixation classification in complex 3D environments remains an important direction for future research.

### 2.2.3 Data Visualization

Data visualization is essential for qualitative analysis of attention in eye tracking studies. The VISTA platform's visualization module allows the inspection session to be replayed from the inspector's point of view with an overlaid gaze position marker. It also supports visualization of the inspector's movement path, the distribution of 3D gaze points, and surface-based attention heatmap on the bridge model.

Two heatmap-based visualizations were generated using external Python tools: a surface-based attention heatmap and an area-revisit heatmap. The surface-based attention heatmap follows prior surface-mapping approaches for eye-tracking visualization [33–35], For each vertex location, $\boldsymbol{g}_i$, on the bridge model, an attention weight, $w_j$, was calculated as a function of its distance from all valid gaze intersection points, $\boldsymbol{p}_j$

$$w_j = \frac{1}{\sigma\sqrt{2\pi}} \sum_{i=1}^{n} exp\left(\frac{-\left\|\boldsymbol{p}_j - \boldsymbol{g}_i\right\|^2}{2\sigma^2}\right), \text{ for } j \in \mathcal{P}, \tag{3}$$

where $n$ is the total number of gaze intersection points, $\sigma$ is a positive-valued smoothing factor controlling the spatial extent of each gaze point. Larger values of $\sigma$ distribute attention over a broader surface region, while smaller values produce more localized attention maps [35]. The resulting vertex weights are calculated for all the vertices, $j$, in the set of surface vertices $\mathcal{P}$, and are normalized using min-max normalization, yielding a surface distribution of visual attention across the bridge model.

In addition to the surface attention heatmap, this study introduces an area revisit heatmap to quantify repeated spatial coverage during inspection. While the attention heatmap identifies where gaze was concentrated, the area-revisit heatmap captures how often regions of the bridge were revisited after a period of absence. This descriptor is intended to reflect inspection redundancy, repeated checking, or inefficient re-examination of previously viewed areas.

To calculate the area viewed by the inspector at a given time, a fixed-size spotlight model was used. Specifically, a 20° cone was projected from the estimated head position in the viewing direction. The portion of the bridge surface intersected by this cone was treated as the viewed area at that time step. Figure 4 illustrates this projection-based calculation for a three-second inspection window, where the viewing cones are cast from successive camera positions to identify the corresponding visible regions on the bridge surface. This approximation provides a consistent basis for comparing spatial coverage across inspectors, although prior research has shown that the spatial extent of attention can vary with task demands [36–38].

An area was counted as revisited when it was viewed again after not being viewed during the preceding 20 seconds. This threshold was selected to distinguish meaningful revisits from continuous viewing or short local adjustments in gaze direction. As the viewing threshold approaches zero, the area revisit heatmap becomes increasingly similar to the fixation heatmap because nearly all repeated viewing contributes to the revisit count. On the other hand, as the viewing threshold becomes very large, the area revisit heatmap will approach a uniform distribution where all areas are visited once because few regions qualify as revisits. A visualization of these limiting behaviors and discussion on the 20 second threshold is shown in Fig. A2 in the appendix.

The area revisit heatmap differs from the surface-based attention heatmap in two important ways. First, the attention heatmap is computed from gaze intersection points, which may be null when the gaze ray does not intersect the model. The revisit heatmap, in contrast, is based on viewing direction and the projected attention cone; therefore, it can still identify viewed bridge regions even when the precise gaze intersection point is missing, erroneous, or noisy. Second, the revisit heatmap emphasizes temporal return behavior rather than cumulative visual attention. This makes it useful for characterizing inspection redundancy and spatial coverage patterns that may not be apparent from gaze concentration alone.

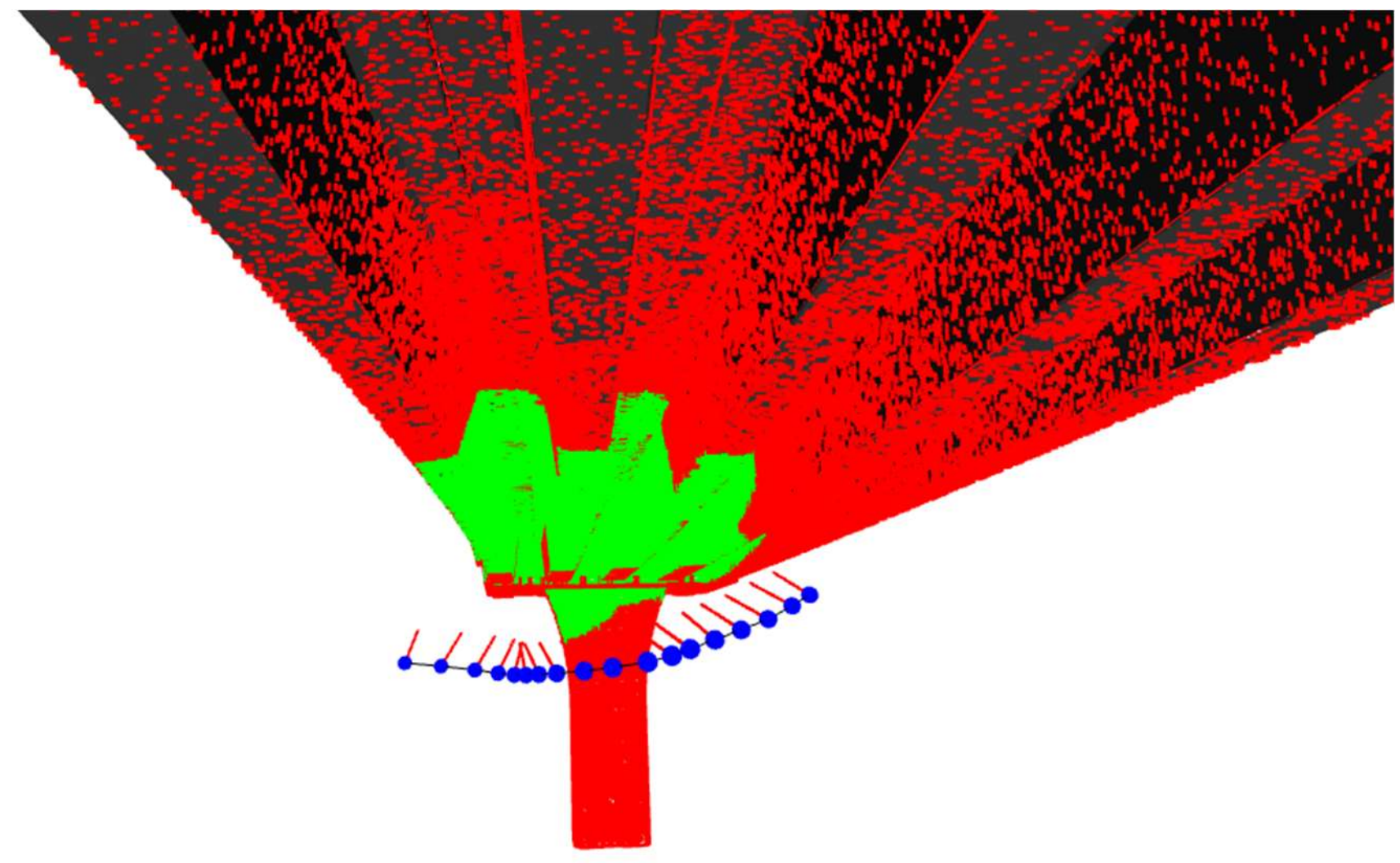

***Fig. 4.*** *Projection-based estimation of viewed frustum with origin at the camera positions (blue spheres) and in the direction of the inspector's gaze (red lines).*

### *2.2.4 Window-based feature extraction*

The proposed pipeline extracts features from short time windows of the inspection data. Each window is represented by gaze, fixation, head movement, and navigation features, which are later used to classify the corresponding time segment as global scanning, local inspection, or navigation. Table 1 lists the features extracted from each window, along with their description. These features were selected based on prior eye-tracking and task-prediction studies that have used gaze, fixation, saccade, entropy, and movement measures to characterize attention allocation, task state, and performance during visual search and inspection tasks [2,8,14,15,19,21,28]. The features were organized into five groups to reflect the different aspects of behavior captured within each temporal window. Viewpoint-control features describe changes in the inspector's viewing position and direction, including head and drone movement, and are intended to capture the navigation and positioning demands of the task. Gaze-response features include continuous eye-tracking measurements, such as pupil diameter and angular eye-movement velocity and acceleration, which describe immediate gaze-related responses during inspection. Fixation–saccade features summarize event-level eye-movement behavior, including fixation frequency, fixation duration, saccade characteristics, and the proportion of time spent fixating. Gaze-distribution features characterize how gaze is spatially organized within the screen during a window, using measures such as fixation dispersion and gaze entropy. Finally, geometry features describe the spatial relationship between the inspector's viewpoint and the bridge model, including distance to the bridge and the visible or inspected bridge area. This grouping separates viewpoint control, eye-movement behavior, spatial gaze organization, and scene context while keeping the features compatible with unconstrained 3D inspection environments.

***Table 1.*** *List of features derived from window data and summary statistics.*

| *Group* | *Feature Name (Unit)* | *Description* | *Ref.* |
|---|---|---|---|
| *Viewpoint Control Features* | *Head Linear Velocity (m/s)* | Rate of change in estimated head position | [39] |
| | *Head Linear Acceleration (m/s)* | Change in head linear velocity | [39] |
| | *Drone Linear Velocity (m/s)* | Translational velocity of the drone camera | [39] |
| | *Drone Turning Velocity (rad/s)* | Angular velocity of the drone camera | [39] |
| *Gaze Response Features* | *Pupil Diameter (mm)* | Mean pupil diameter within the window | [12,15] |
| | *Eye Movement Velocity (deg/s)* | Angular gaze velocity | [2] |
| | *Eye Movement Acceleration (deg/s$^2$)* | Change in angular gaze velocity | [2] |
| *Fixation-Saccade Features* | *Number of Fixations (ea.)* | Count of fixations detected within window | [15] |
| | *Fixation Duration (s)* | Duration of detected fixations | [15,21] |
| | *Saccade Velocity (deg/s)* | Velocity during saccadic movement | [15,21] |
| | *Saccade Amplitude (deg)* | Angular distance covered by saccades | [15,21] |
| | *% Time in Fixation* | Fraction of window classified as fixation | N/A |
| *Gaze Distribution Features* | *Spatial Variance of Fixations (deg)* | Spatial spread of fixation points | [21] |
| | *Stationary Gaze Entropy [SGE]* | Distribution of gaze across screen regions | [28] |
| | *Transition Gaze Entropy [GTE]* | Transitions of gaze between regions | [28] |
| *Scene Geometry Features* | *Distance to Bridge (m)* | Distance from viewpoint to bridge surface | [39] |
| | *Area of Bridge Inspected (%)* | Surface area viewed within the window | N/A |

The I-VT fixation filter was used to calculate fixation-based metrics. In addition to fixation measures, stationary gaze entropy and transition gaze entropy were calculated following the methodology summarized by Shiferaw [28]. These entropy measures characterize the spatial distribution and transition structure of gaze across screen regions within each temporal window. For entropy calculation, the screen was divided into a nine-panel layout, with the center panel subdivided into nine smaller regions. This design reflects the tendency for gaze points to cluster near the center of the screen, which is typically the most comfortable viewing position and corresponds to the drone camera’s forward direction. Subdividing the center region provided finer resolution for distinguishing gaze behavior that remained within the central field of view while avoiding excessive sparsity in peripheral regions.

The distinction between window-based features and aggregate descriptors is important in this study. Window-based features describe short segments of the inspection and are used as inputs to the inspection-mode classifier. Aggregate descriptors, by contrast, summarize the entire inspection session after the inspection mode time series has been generated. This separation allows low-level gaze and movement measurements to be used for behavioral segmentation, while higher-level temporal descriptors are used to characterize inspection strategy and evaluate performance.

### 2.3 Data collection

Data collection was conducted in VISTA to obtain gaze, movement, and performance data during a simulated bridge inspection task. A total of eight users were recruited and asked to inspect a three-span concrete highway bridge for visible surface defects. It should be noted that the data from the small user study aims to demonstrate the feasibility and applicability of the system rather than to produce conclusive, generalizable insights. Figure 5(a) shows the experimental setup and an example participant’s view during inspection.

The inspection environment was created by adding 24 synthetic defects to a 3D concrete bridge model using the Blender computer graphics package [40]. The defects were distributed across the three central spans of the bridge and represented visible surface deterioration. Examples of the defects are shown in Figure 5(b). Before beginning the inspection, participants completed a tutorial to become familiar with the drone navigation controls. During the inspection task, they were given 15 minutes to inspect the bridge and identify as many defects as possible. After the inspection task, participants completed a post-study survey documenting their prior experience with bridge inspections, video games, structural engineering, and drone operation. Participants also completed the Purdue Spatial Visualization Test: Rotations (PSVT:R) [41], which was used as a measure of spatial ability. Spatial visualization was included because it has been shown to correlate with performance in various engineering tasks [42] and may be relevant to interpreting inspection behavior in a navigable 3D environment.

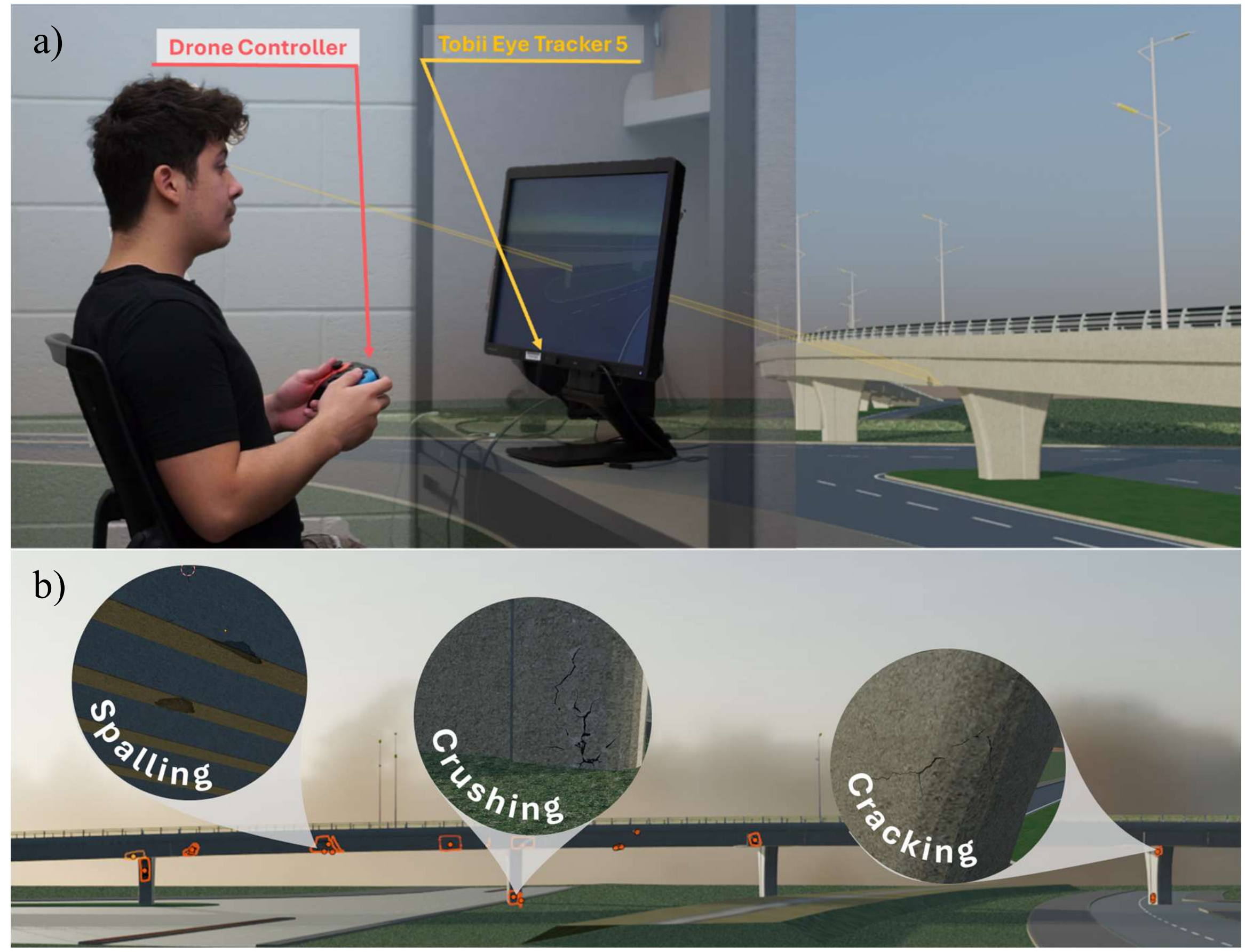

***Fig. 5.*** *User study environment setup: (a) Example participant's view. (b) Synthetic defect examples.*

## 2.4 Inspection mode labeling

Supervised learning was used to classify the inspection windows into inspection modes. To obtain training labels, a group composed of four labelers reviewed replay videos from the eight inspection sessions. The labelers were asked to assign an inspection mode to randomly selected three-second windows throughout each inspection. Before labeling, all labelers were provided with the definitions of each inspection mode. Additionally, a fourth option, disengaged, was also included for periods in which the participants appeared distracted by something outside the inspection environment or were otherwise not engaged in visual search, assessment, or navigation. Windows labeled as disengaged were excluded from model training to avoid introducing ambiguous or non-task-related behavior into the classifier.

During each replay, a red border appeared around the video three seconds before the labeling time to direct the labelers' attention to the relevant window. After the three-second interval, the labelers were asked: *"What was the primary inspection mode during the time window?"* If more than one inspection mode occurred within the window, such as during a transition between navigation and local inspection, they were instructed to choose the mode that occupied the largest portion of the window. Labelers were allowed to replay the video, view the surrounding context, and discuss the window until they reached a consensus label. They assigned a confidence level to each label using three categories: low, medium, or high.

A consensus-based labeler group format was used because preliminary labeling trials showed that disagreement was primarily due to differences in how labelers interpreted the boundaries between inspection modes. Group discussion allowed the labelers to refine these interpretations and apply a more consistent definition across participants and inspection sessions. The authors served as moderators to explain the labeling procedure, clarify mode definitions when asked, and monitor labeler fatigue during the session. To reduce bias in the labeling process, the moderators did not participate in assigning labels or resolving disagreements.

## 2.5 Inspection mode prediction

After obtaining labels, each labeled sample was paired with the three-second window of raw data immediately preceding the label time. Each window was processed into a set of 15 window-based features for inspection mode classification. The features and their summary statistics are shown in Table 1. Four machine learning models were evaluated: Artificial Neural Networks (ANN) [43], Random Forest [44], Support Vector Machines (SVM) [45,46], and Extreme Gradient Boosted Trees (XGBoost) [47]. These models were selected to compare commonly used classifiers with different levels of complexity, nonlinearity, and interpretability. Per-user feature normalization was applied to reduce differences in baseline gaze and movement behavior across participants. Sample weights based on label confidence were also incorporated during training, assigning greater influence to windows with higher confidence consensus labels.

Model development used both 10-fold cross-validation and leave-one-user-out (LOUO) cross-validation. In 10-fold cross-validation, labeled windows were divided into training and testing folds across the full dataset. In LOUO cross-validation, models were trained on seven users and tested on the remaining participant. LOUO validation was treated as the primary criterion for hyperparameter selection because it evaluates generalization to unseen users, which is more representative of the intended application than random window-level splits.

During inference, each user's inspection was divided into three-second windows with 2.5 seconds of overlap. Each trained model produced a class probability vector for every window. The final label prediction for each window was obtained by averaging the class probabilities from the four classifiers and selecting the class with the highest mean probability:

$$\hat{y}_j = \underset{k\in\{1,2,3\}}{\operatorname{argmax}} \frac{1}{F}\sum_{i=1}^{F} f_{ik}(x_j)\,, \tag{4}$$

where $\hat{y}_j$ is the final prediction corresponding to window $j$, $f_{ik}$ is the probability assigned by classifier $i$ to class $k$, and $x_j$ is the feature vector for window $j$.

Since the inference windows overlapped, each time point was covered by multiple window predictions. To select a single label at time $t$, predictions from all windows containing that time point were combined using a distance-weighted vote. Windows with centers closer to $t$ were assigned greater weight:

$$\hat{y}_t = \underset{k\in\{1,2,3\}}{\operatorname{argmax}} \sum_{j\in\mathcal{W}(t)} \frac{1}{|t - t_j|}\mathbb{I}(\hat{y}_j = k) \tag{5}$$

where $\hat{y}_t$ is the predicted inspection mode at time $t$, $\mathcal{W}(t)$ is the set of windows containing $t$, $t_j$ is the center time of window $j$, and $\mathbb{I}(\hat{y}_j = k)$ is an indicator function equal to 1 when window $j$ is assigned to class $k$. This procedure produces a continuous inspection mode time series for each participant.

### 2.6 Inspection summary and correlation analysis

The proposed procedure yields an inspection mode time series that encodes the participant's behavioral state at every point in time. This time series provides a compact representation of how the participant transitions among global scanning, local inspection, and navigation. Signal processing techniques and summary statistics were then applied to this signal to extract aggregate descriptors of inspection behavior.

Three groups of aggregate descriptors were obtained. The first group was directly derived from the inspection mode time series and included measures related to the frequency, duration, and transition structure of the predicted inspection modes. These descriptors were intended to characterize the inspection strategy at the behavioral level. The second group was derived from raw fixation data, such as the number of fixations and the fixation-to-saccade ratio, which have been used in previous studies [2,48]. The third group was calculated based on the inspectors' area revisit data. The descriptor definitions are provided in Table 2.

***Table 2.*** *Definition of aggregate descriptors.*

| ***Source*** | ***Descriptor*** | ***Definition*** |
|---|---|---|
| Inspection mode time series | *Time in navigation* | Percentage of inspection classified as navigation. |
| | *Time in global scanning* | Percentage of inspection classified as global scanning. |
| | *Time in local inspection* | Percentage of inspection classified as local inspection. |
| | *Local to global ratio* | The ratio of total time spent in local inspection over global scanning. |
| | *Navigation dwell time* | The average consecutive time spent on navigation. |
| | *Global scanning dwell time* | The average consecutive time spent on global scanning. |
| | *Local inspection dwell time* | The average consecutive time spent on local inspection. |
| Inspection mode transition | *Navigation to local TP* | Probability of transitioning from navigation to local inspection mode. |
| | *Global scanning to local TP* | Probability of transitioning from global scanning to local inspection mode. |
| | *Local to global TP* | Probability of transitioning from local inspection to global scanning mode. |
| | *Global to navigation TP* | Probability of transitioning from global scanning to navigation mode. |
| | *Inspection mode transition entropy* | Conditional entropy of inspection mode transition probabilities. |
| *Redundancy Heatmap* | *Mean redundancy factor* | The average number of times any point of the bridge is viewed. |
| | *Redundant areas* | Percentage of points on the bridge that are viewed more than once. |
| *Raw gaze data* | *Total number of fixations* | Number of fixations in the entire inspection period. |
| | *Fixations to Saccade ratio* | Ratio of time spent in fixating versus time spent in saccadic movement. |

A descriptor-metric correlation analysis was then performed to identify which behavioral descriptors were associated with inspection performance. This analysis was intended as a feasibility assessment of whether the proposed inspection mode time series and area-revisit descriptors can provide meaningful indicators of inspection skill and performance in unconstrained 3D bridge inspection.

# 3 Results and Discussion

## 3.1 User study and survey summary

Figure 6 summarizes the participant background, spatial ability scores, inspection coverage, and defect detection outcomes for the eight participants, which shows a large variability in the self-reported familiarity with drone operation, video games, and bridge inspections, indicating meaningful differences in skills that may affect navigation, visual search, and adaptation to the VISTA platform. However, their familiarity with structural engineering concepts showed lower variability since the participant pool consisted of graduate students in structural engineering. The percentage area of the bridge area inspected ranged from 58% to 85%, indicating that most participants covered a comparable portion of the inspection environment within the 15-minute time limit. This relatively narrow range suggests that participants may have adopted similar implicit criteria for what constituted sufficient inspection coverage. However, similar spatial coverage did not necessarily imply similar inspection performance, because participants differed in the number of defects documented, the time required to identify defects, and the degree of repeated inspection.

Defect detection was evaluated using two complementary criteria: defects identified via images and defects identified via gaze. A defect identified through image capture was considered explicitly documented by the inspector. A defect identified through gaze was defined as a defect that appears to have been visually observed based on eye-tracking data but was not necessarily documented with an image. This distinction is important because an inspector may look directly at a defect without recognizing it as damage, may recognize it but fail to capture an image, or may delay image capture while navigating to a better viewpoint.

Across participants, the number of defects identified via images and gaze ranged from 13 to 18 and 17 to 23, respectively. The higher gaze-based counts indicate that some defects were visually encountered but not documented. Upon further inspection, it was found that none of the users captured images of the hairline cracks on the column shown on the right in Figure 5(b), whereas all users identified the more visually salient scouring and spalling defects shown on the left in Figure 5(b). This outcome suggests that the task contained defects of varying difficulty. At the same time, the relatively low variability in the number of documented defects indicates that many defects were visually salient enough to be detected by most participants.

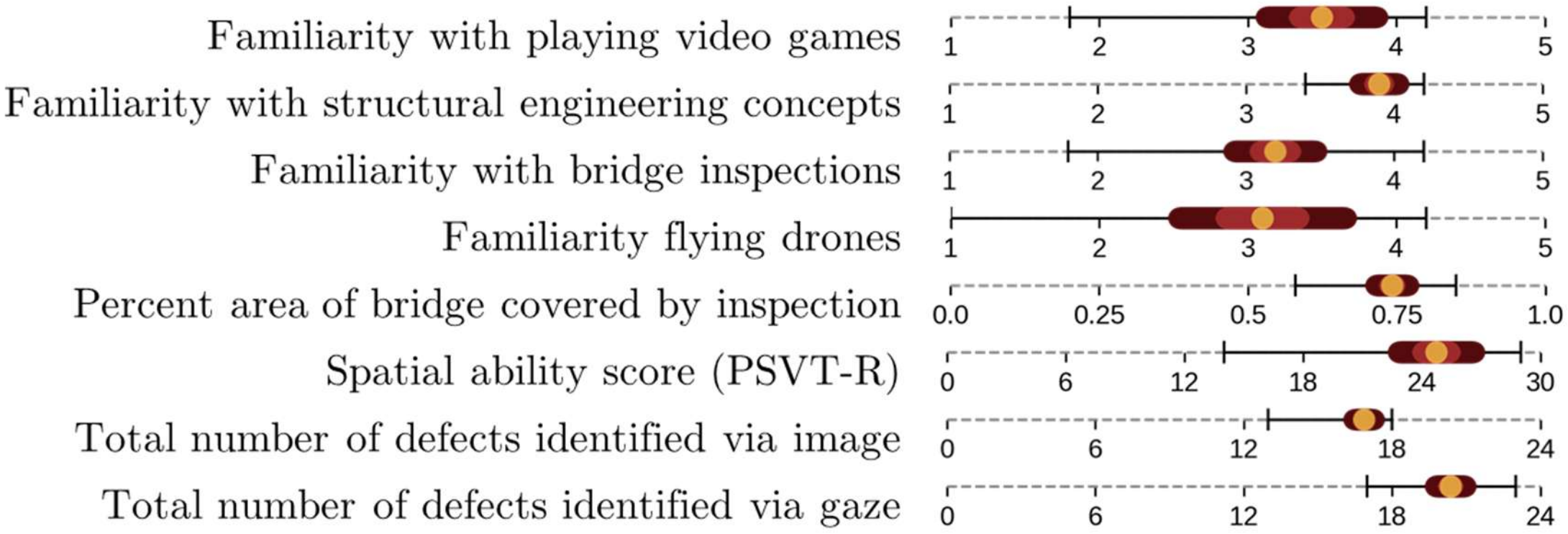

***Fig. 6.*** *Summary statistics of bridge inspection performance metrics, spatial ability scores, and survey responses.*

### 3.2 Qualitative inspection analysis

VISTA visualizations were first used to support qualitative interpretation of inspection behavior and to determine whether defects were visually observed or explicitly documented. The criteria for determining whether a defect was found using gaze is further explained in the appendix.

The gaze heatmap and area revisit heatmaps were calculated to compare spatial attention patterns across inspectors. Figure 7 compares users 3 and 5. In the surface-based gaze heatmap, user 5 exhibits several localized regions of elevated gaze salience across the bridge, suggesting focused attention on specific structural regions or candidate defects. User 3, in contrast, shows a more diffused gaze distribution, with fewer sharply localized high-salience areas. This pattern suggests that user 3 distributed visual attention more broadly across the structure rather than concentrating on a smaller number of specific regions. The area-revisit heatmap in Figure 9 adds temporal context to this comparison. User 5 shows more high-revisit regions than user 3, indicating more frequent returns to previously viewed areas. This repeated and redundant coverage may reflect a more thorough inspection strategy, in which the inspector revisited regions to verify observations, improve viewing angle, or confirm potential defects before documentation. The area revisit heatmap offers a possible explanation for the gaze heatmap, since user 5 has a higher redundancy distributed throughout the bridge, it is likely that this user went back to check for missing defects around their already found salient areas, leading to more localized saliency around some defects.

These qualitative differences are consistent with differences in inspection outcomes. User 3 documented 13 defects, whereas user 5 captured 18 defects. The difference in performance can not be attributed to a single factor, because inspection performance may depend on spatial ability, navigation skill, search strategy, and defect recognition. For example, User 5 had a higher PSVT:R score than User 3. Nevertheless, the heatmaps show that VISTA can reveal differences in how inspectors distribute attention and revisit regions, which provides a useful basis for interpreting performance differences beyond defect counts alone.

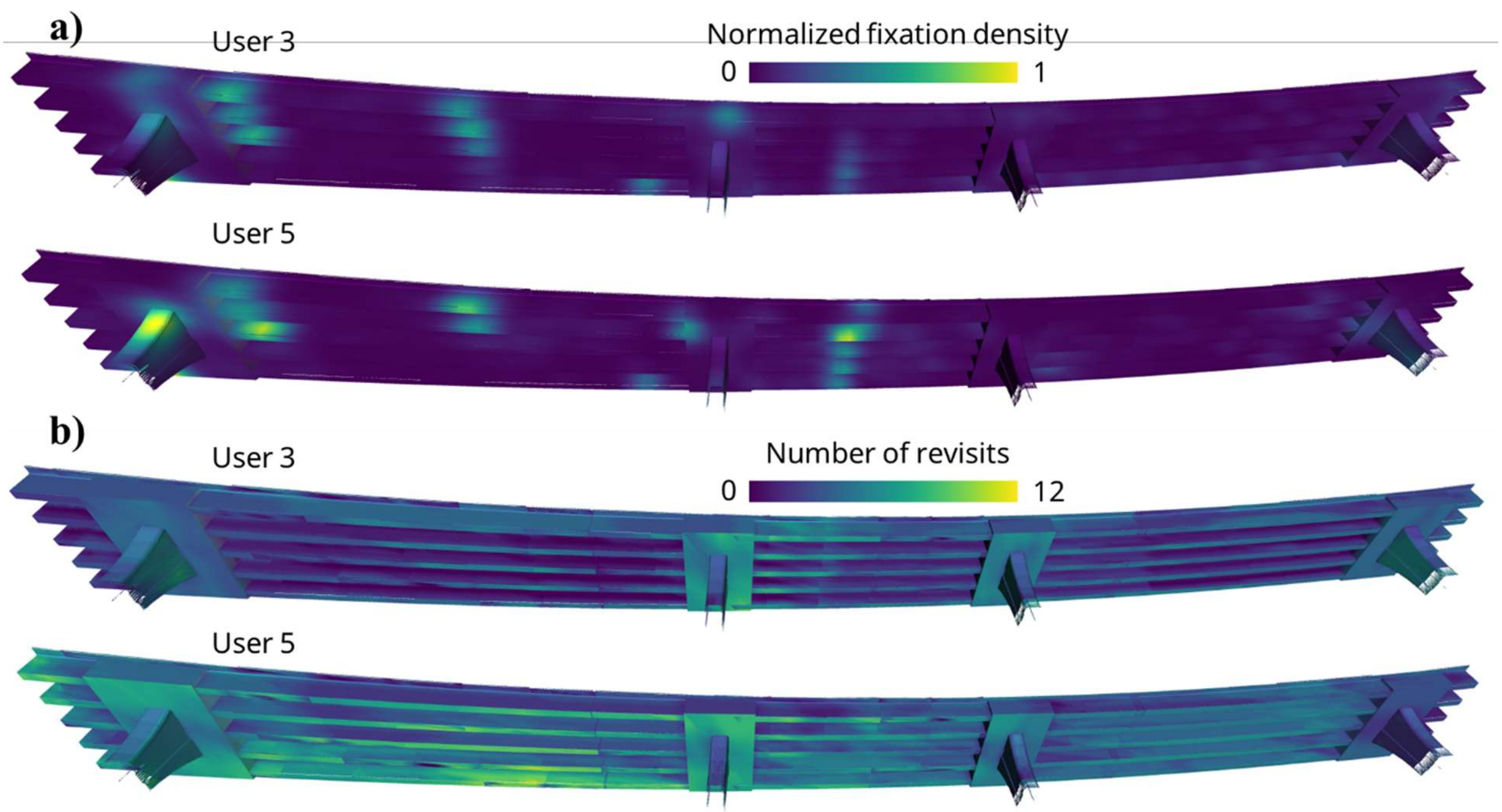

***Fig.** 7. Qualitative heatmap comparison between two user inspections. (a) Gaze heatmap. (b) Area revisit heatmap.*

### 3.3 Inspection mode labeling and classifier performance

A total of 20-50 random label times were selected for each inspector based on inspection duration. Figure 8 shows the distribution of global scanning, local inspection, and navigation labels for all eight users. After removing the samples labeled as disengaged, the final labeled dataset contained 310 samples. Labeling was completed over three sessions, each lasting 2–4 hours, with breaks included to reduce fatigue and maintain labeling quality.

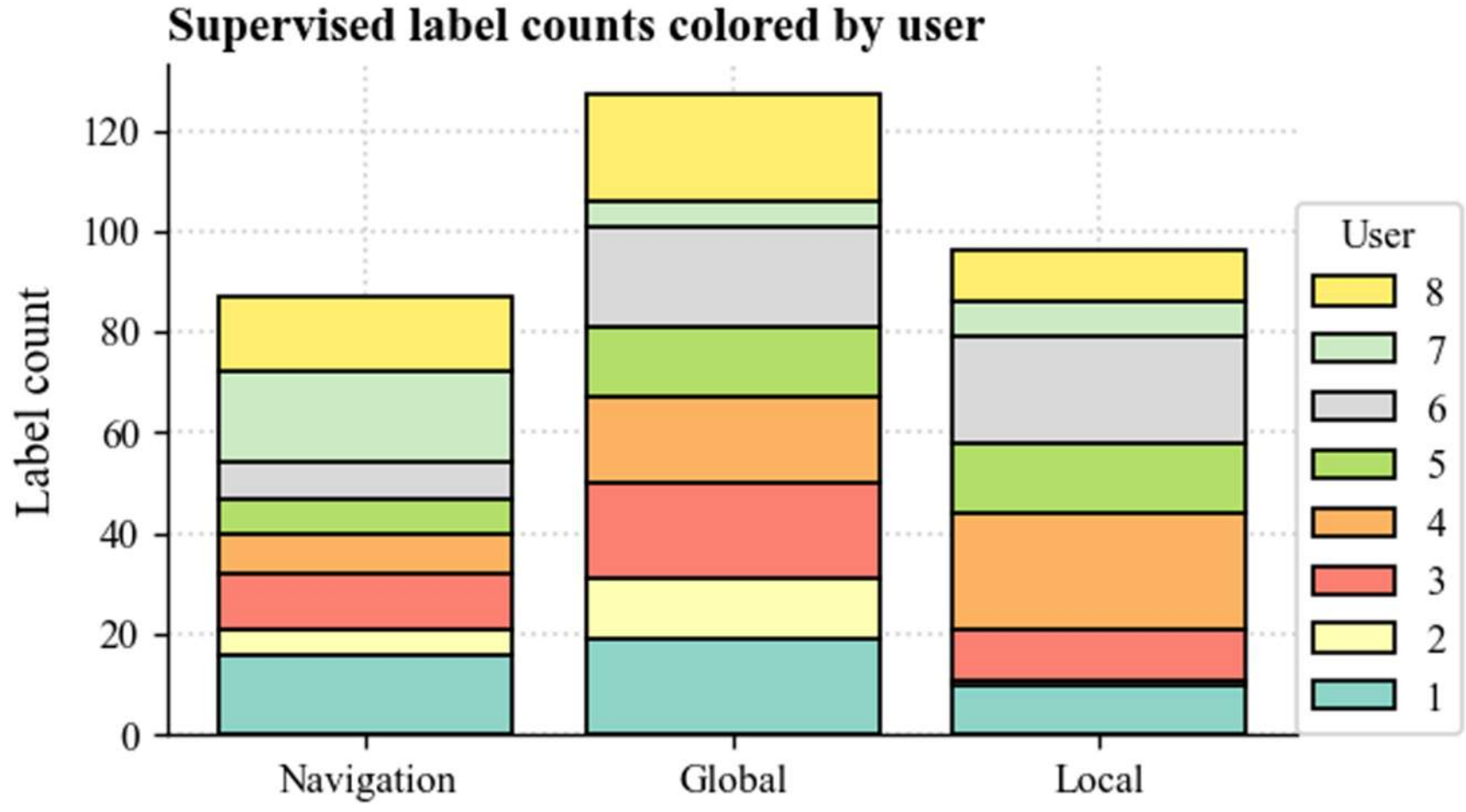

***Fig. 8.*** *Distribution of inspection mode labels for all users.*

An important observation from the labeling sessions was that the operational definitions of the inspection modes became more consistent as the labeling progressed. Early in the process, labelers required more discussion to distinguish between modes, particularly during transitions. Later in the process, the group reached consensus more efficiently. This observation suggests that inspection modes are interpretable but require calibration among labelers. The labeling group format was therefore useful for producing a consistent labeled dataset, although it also highlights the subjectivity inherent in supervised behavioral labeling.

Before evaluating classifier performance, the labeled windows were examined to determine whether the selected features exhibited interpretable differences across inspection modes. Figure 9 shows the mean normalized feature vector for windows labeled as navigation, global scanning, and local inspection. The observed patterns are consistent with the intended definitions of the modes: navigation is characterized by stronger viewpoint-control features, local inspection is associated with lower movement and lower gaze entropy, and global scanning shows greater gaze dispersion and entropy. These differences provide empirical support for using the selected features for inspection mode classification.

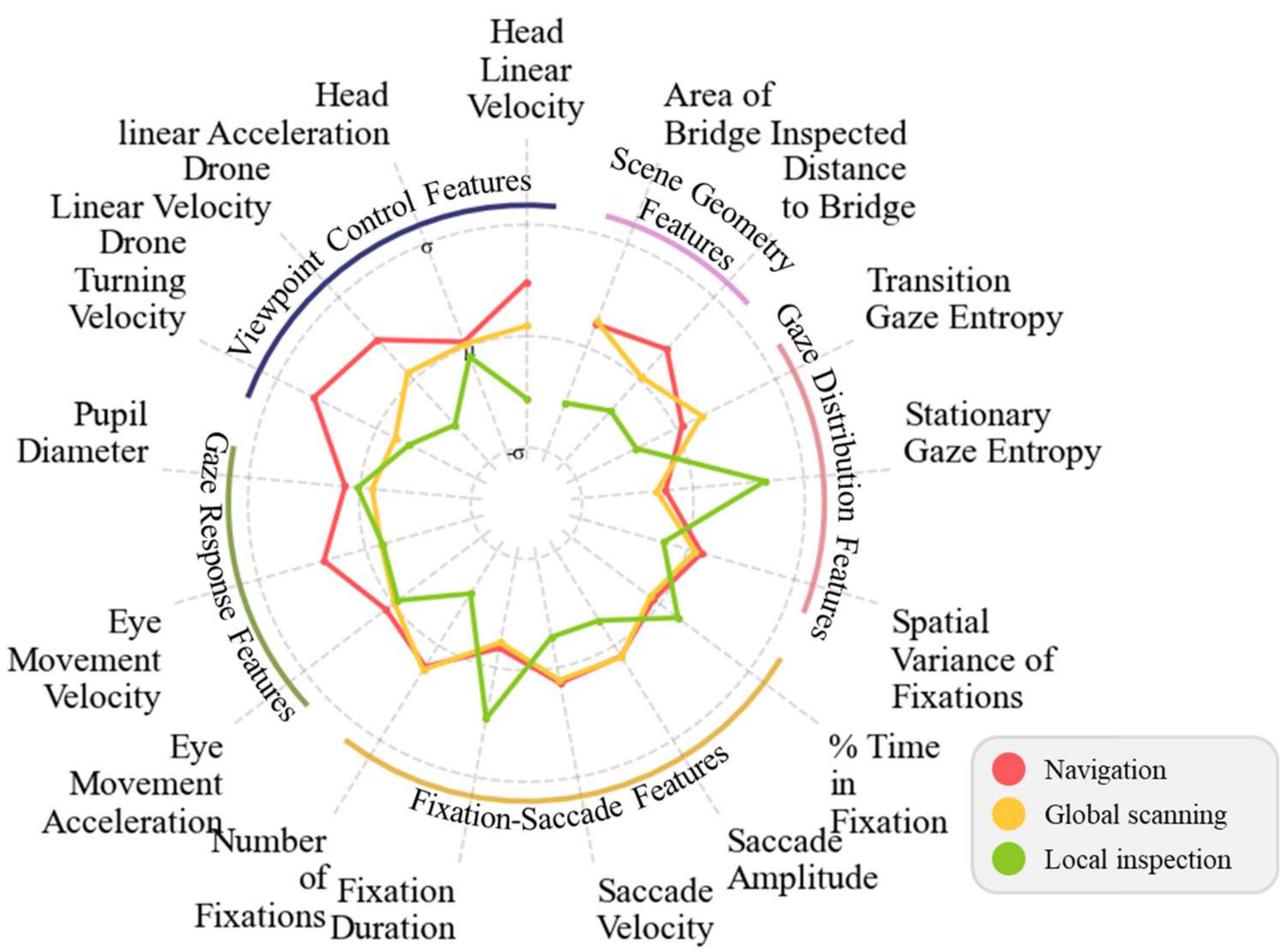

***Fig. 9.*** *Comparison of mean feature behavior during windows labeled with different inspection modes.*

Table 3 reports the three-trial mean model performance for the best set of hyperparameters found using grid-search. The near-perfect 10-fold cross-validation scores indicate that the models can learn the labeled patterns when training and testing windows are randomly drawn from the same participant pool. However, performance decreased in leave-one-user-out cross-validation, indicating that generalization across users is significantly more difficult. This result is expected due to the inherent variability in how inspectors express the same inspection modes through gaze, movement, and navigation. For this reason, leave-one-user-out (LOUO) performance was treated as the primary evaluation criterion. Based on the LOUO performance, it can be seen that the model ensemble based on model confidence voting improves LOUO generalization performance.

***Table 3.*** *Classification performance under 10-fold and LOUO cross-validation.*

| **Model** | **10-Fold Cross-Validation** | | | **LOUO Cross-Validation** | | |
|---|---|---|---|---|---|---|
| | *Mean Train Acc.* | *Mean Test Acc.* | *Std. Test Acc.* | *Mean Train Acc.* | *Mean Test Acc.* | *Std. Test Acc.* |
| ANN | 0.95 | 0.94 | 0.02 | 0.96 | 0.67 | 0.11 |

| Random Forest | 0.98 | 0.95 | 0.02 | 0.98 | 0.69 | 0.11 |
|---|---|---|---|---|---|---|
| SVM | 0.91 | 0.89 | 0.03 | 0.92 | 0.68 | 0.09 |
| XGBoost | 1.00 | **0.99** | 0.01 | 1.00 | 0.67 | 0.12 |
| Model Ensemble | 0.97 | 0.96 | 0.02 | 0.98 | **0.72** | 0.11 |

To evaluate the effect of window size on model training results, an ablation study was performed by training the models with the best hyperparameters while varying the window sizes. Table 4 reports the average cross-validation performance across 3 trials for the model ensemble trained using all input features. The ablation study shows that a 3-second window size improved generalization under LOUO validation, while all window sizes performed well in 10-fold cross-validation. This suggests that longer windows may provide a more stable behavioral context for distinguishing inspection modes across users. However, larger windows also reduce temporal resolution, eventually leading to degraded performance. The selected three-second window, therefore, balances classification performance with the need to capture short behavioral transitions.

***Table 4.*** *Ablation study on the effect of window size on cross-validation results.*

| | **10-Fold Cross-Validation** | | **LOUO Cross-Validation** | |
|---|---|---|---|---|
| **Window Size** | *Mean Test Acc.* | *Std. Test Acc.* | *Mean Test Acc.* | *Std. Test Acc.* |
| 0.5 s | 0.94 | 0.02 | 0.61 | 0.12 |
| 1.0 s | 0.95 | 0.02 | 0.64 | 0.09 |
| 3.0 s | **0.96** | 0.02 | **0.71** | 0.13 |
| 4.0 s | **0.96** | 0.02 | 0.65 | 0.14 |

Table 5 evaluates the effect of feature groups on cross-validation performance using a 3-second window size. The best performance was obtained when all feature groups were included. However, gaze response features only slightly improved 10-fold test accuracy, and could be omitted without degrading LOUO accuracy. Viewpoint control features contributed most strongly to classification, achieving over 60% LOUO accuracy on their own. While gaze distribution features seem to be the most important attention-related features, they are not sufficient for classification on their own.

Scene geometry features provided similar gains to gaze distribution features, suggesting that they may be useful but not essential in all applications. This distinction is important for future deployment outside virtual environments, where geometric features may be more difficult to obtain than gaze distribution and viewpoint control features.

***Table 5.*** *Ablation study on the effect of feature set on cross-validation results.*

| Feature Set | 10-Fold | | LOUO | |
|---|---|---|---|---|
| | *Mean Test Acc.* | *Std. Test Acc.* | *Mean Test Acc.* | *Std. Test Acc* |
| All features | 0.96 | 0.02 | 0.71 | 0.11 |
| Without viewpoint control features | 0.96 | 0.02 | 0.52 | 0.09 |
| Without gaze response features | 0.95 | 0.02 | 0.71 | 0.11 |
| Without fixation-saccade features | 0.95 | 0.02 | 0.70 | 0.13 |
| Without gaze distribution features | 0.97 | 0.02 | 0.69 | 0.13 |
| Without scene geometry features | 0.96 | 0.02 | 0.67 | 0.13 |
| Viewpoint control features only | 0.87 | 0.04 | 0.67 | 0.10 |
| Gaze distribution features only | 0.78 | 0.04 | 0.52 | 0.12 |
| Scene geometry features only | 0.74 | 0.03 | 0.48 | 0.16 |
| Viewpoint control + gaze distribution features | 0.90 | 0.03 | 0.67 | 0.12 |
| Viewpoint control + scene geometry features | 0.92 | 0.03 | 0.68 | 0.12 |

Table 6 reports inter-model agreement on cross-validation test-set predictions using Krippendorff's Alpha. Based on this table, agreement was highest for inter-user prediction when using viewpoint control features alone and dropped when additional feature groups were included. One possible explanation is that adding attention and geometry features increases model flexibility, allowing different classifiers to use different decision boundaries. Although this reduced agreement over all windows, the full feature set improved generalization performance. On the other hand, model agreement improves with more feature groups for 10-fold cross-validation predictions. This distinction happens because all four models are able to achieve near perfect 10-fold cross-validation performance, therefore converging close to the optimal decision boundary.

***Table 6.*** *Ablation study on the effect of feature set on inter-model agreement.*

| **Feature Set** | **10-fold Krippendorff's Alpha** | **LOUO Krippendorff's Alpha** |
|---|---|---|
| All features | **0.88** | 0.61 |
| Without viewpoint control features | 0.82 | 0.51 |
| Without gaze response features | 0.86 | 0.61 |
| Without fixation-saccade features | 0.87 | 0.63 |
| Without gaze distribution features | 0.87 | 0.59 |
| Without scene geometry features | 0.86 | 0.61 |
| Viewpoint control features only | 0.80 | **0.67** |
| Gaze distribution features only | 0.58 | 0.45 |
| Scene geometry features only | 0.67 | 0.61 |
| Viewpoint control + gaze distribution features | 0.82 | 0.65 |
| Viewpoint control + scene geometry features | 0.82 | 0.58 |

*3.4. Inspection mode time series and behavioral descriptors*

The trained classifiers were applied to each inspection session to generate an inspection mode time series. Figure 10 shows the resulting inspection mode sequence for all eight inspectors, along with summary measures of the percentage of time and average dwell time in each mode. These time series provide a compact representation of how each inspector transitioned among global scanning, local inspection, and navigation during the task. The inspection mode time series reveals qualitative differences in behavioral patterns. For example, new defects were often found during extended periods of local inspection, which is consistent with the interpretation of local inspection as focused assessment of a candidate defect or structural region. Additionally, users 4 and 6 exhibited a similar alternating pattern between local inspection and global scanning, with relatively limited navigation in between. However, user 6 tended to remain in each mode for longer periods (i.e., higher dwell time), suggesting slower but more sustained behavioral phases. These differences are difficult to capture using aggregate fixation counts alone but become visible in the inspection mode time series.

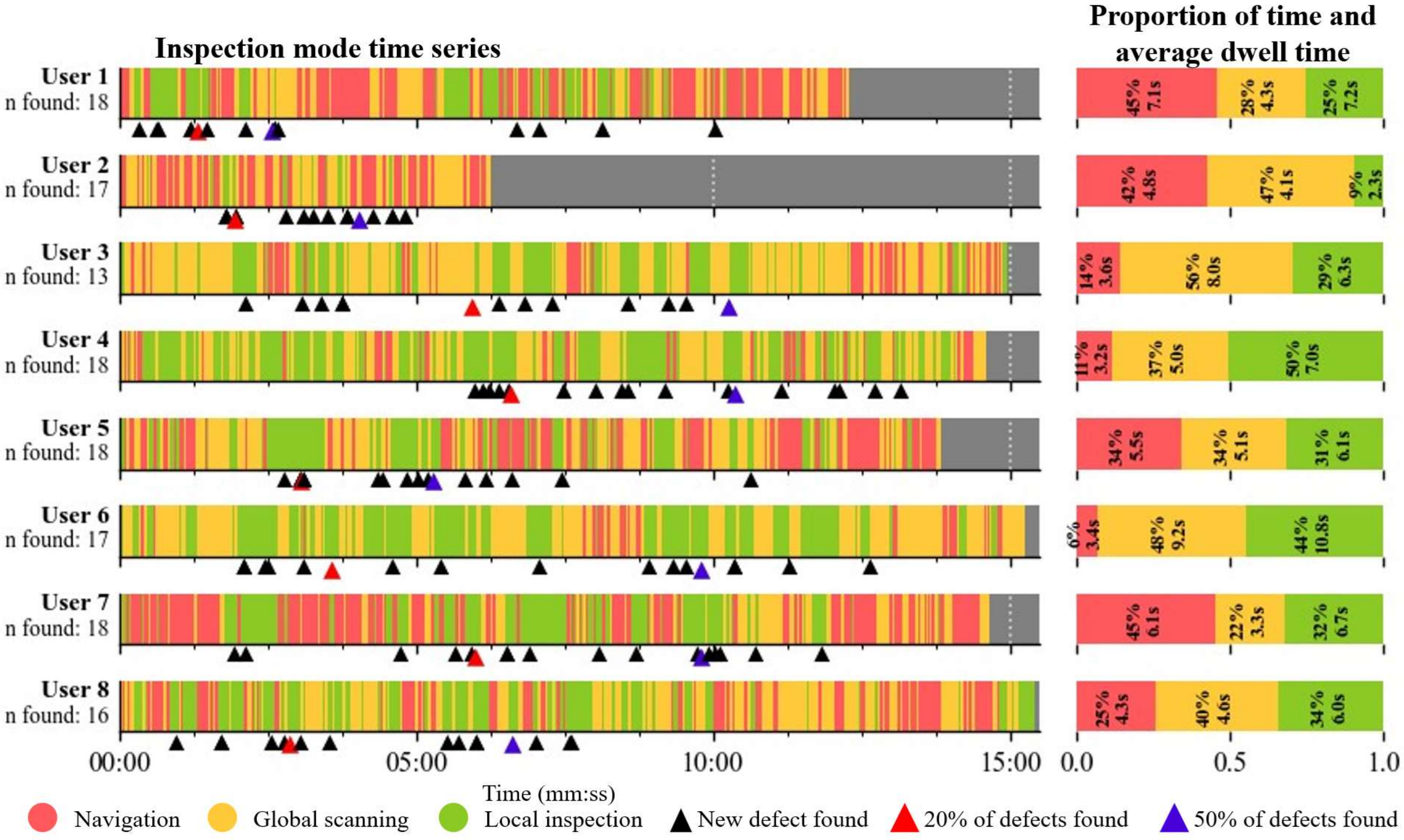

***Fig. 10.*** *Inspection mode time series (left) and percentage of time spent and average dwell time in each cluster (right) for all inspectors.*

The time series also enables transition-based analysis to enable the extraction of quantitative measures of inspector behavioral patterns. Figure 11 shows first-order Markov transition probability matrices for each user's inspection mode time series. These matrices summarize the probability of transitioning from one mode to another and provide a quantitative description of inspection structure. This figure shows that the predicted inspection modes form temporally persistent behavioral states. For all users, self-transition probabilities were substantially higher than cross-mode transition probabilities, with navigation, global scanning, and local inspection self-transitions ranging from 84.4%–93.0%, 85.4%–94.8%, and 78.7%–95.4%, respectively. This indicates that the inspection mode time series captures sustained behavioral phases rather than unstable frame-level predictions. User 2 exhibited the highest transition entropy (2.02) and the lowest local-inspection persistence (78.7%), together with a strong transition from local inspection to global scanning (18.7%), suggesting frequent alternation between focused assessment and broader search. In contrast, User 6 had the lowest entropy (1.33) and the highest persistence in global scanning (94.8%) and local inspection (95.4%), indicating longer uninterrupted behavioral phases. Users 4 and 6 showed relatively low transition probabilities from global scanning to navigation and from local inspection to navigation, suggesting that their inspections were characterized more by alternation between scanning and assessment than by frequent movement-control episodes. Together, these patterns show that transition-based descriptors can capture meaningful differences in inspection rhythm, including behavioral persistence, scan–assess alternation, and reliance on navigation across users.

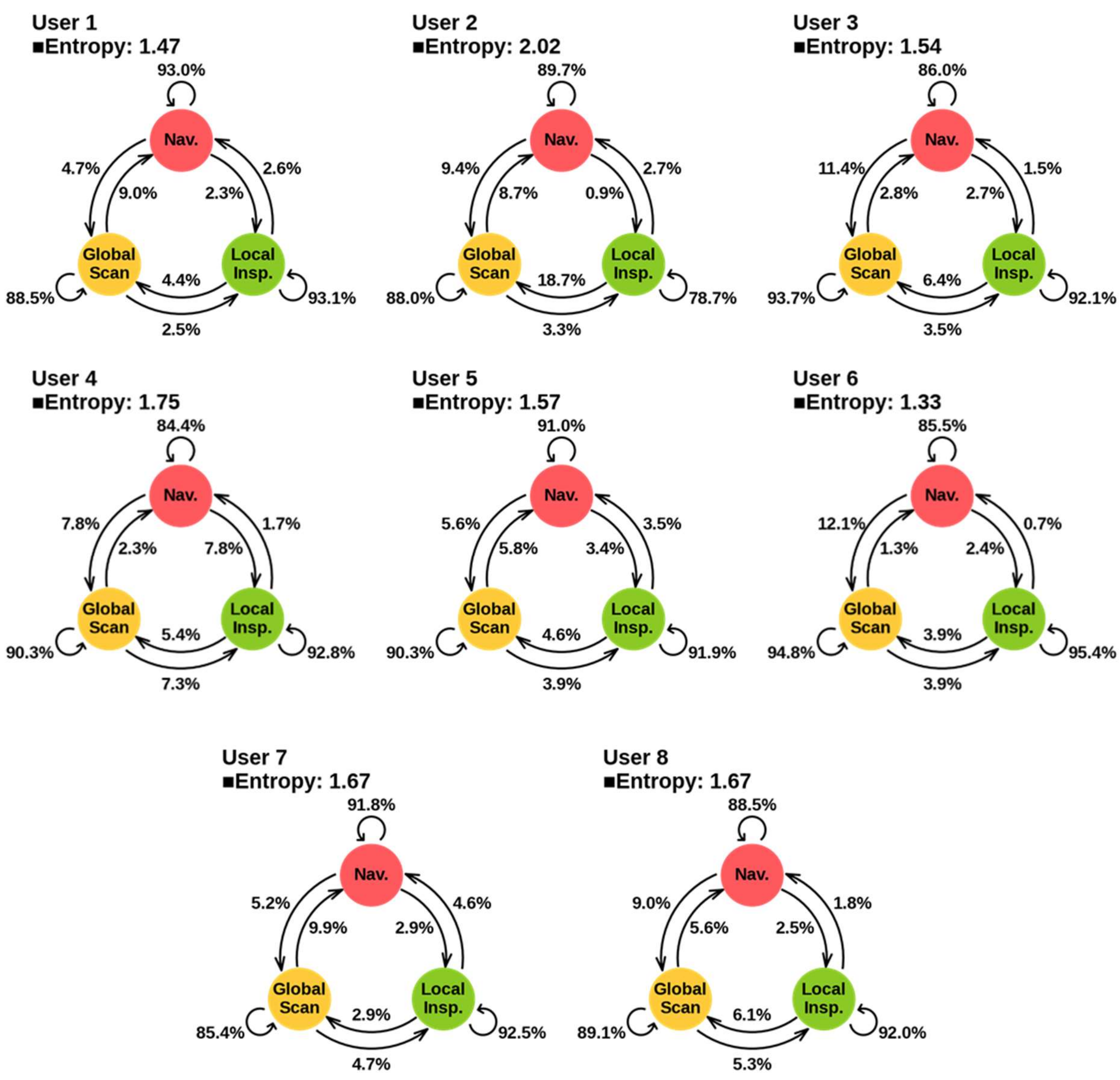

***Fig. 11.*** *Markov transition matrices for the user's inspection mode.*

Table 7 presents the set of aggregate descriptors extracted from the inspection mode time series, raw fixation data, and area-revisit data, along with their summary statistics. A set of nine metrics, summarized in Table 8, was used to evaluate inspection performance in terms of speed, completeness, and spatial coverage. The first set of metrics are related to search efficiency, such as the time spent to find the first 20% or 50% of defects, or the total number of defects found divided by the total inspection time, which we call the discovery rate. These metrics represent how quickly an inspector is able to move through the search process. The second set of metrics is related to the inspectors' visual search completeness, which includes the number of defects seen with gaze and the number of defects found and reported with an image capture. The former measures the degree to which an inspector visually covered the areas of possible defects, and the latter measures the inspector's ability to detect and identify defects within their visual field. The percent bridge area covered is used as a metric because it indicates the inspector's thoroughness in searching beyond just salient regions. Spatial ability score, measured using PSVT:R, is included as a metric since this test has been shown to correlate with performance in a variety of tasks [42]. The inspection aggregate descriptors could then serve as an alternative way to measure this score specific to bridge inspectors.

***Table 7.*** *List of aggregate descriptors derived from user inspections and their summary statistics.*

| Descriptor Name (units) [Short name] | Range | Mean | Std |
|---|---|---|---|
| % time spent on navigation [t nav] | [0.07, 0.46] | 0.282 | 0.160 |
| % time spent in global scanning [t glob] | [0.22, 0.56] | 0.394 | 0.111 |
| % time spent in local inspection [t loc] | [0.10, 0.51] | 0.323 | 0.124 |
| Ratio of time spent in local to global [loc: glob] | [0.20, 1.45] | 0.890 | 0.401 |
| Mean dwell time in navigation (s) [μ nav dwell) | [3.20, 7.10] | 4.772 | 1.400 |
| Mean dwell time in global scanning (s) [μ glob dwell] | [3.31, 9.23] | 5.458 | 2.048 |
| Mean dwell time in local inspection (s) [μ nav dwell] | [2.34, 10.82] | 6.551 | 2.302 |
| TP from navigation to local inspection [nav → loc]* | [0.01, 0.08] | 0.031 | 0.020 |
| TP from global scanning to local inspection [glob → loc]* | [0.03, 0.07] | 0.043 | 0.015 |
| TP from local inspection to global scanning [loc → glob]* | [0.03, 0.19] | 0.066 | 0.050 |
| TP from global scanning to navigation [glob → nav]* | [0.01, 0.10] | 0.057 | 0.033 |
| Inspection mode transition entropy [$H_m$] | [1.33, 2.02] | 1.625 | 0.207 |
| Mean redundancy factor [μ area redundancy] | [2.44, 4.27] | 3.267 | 0.684 |
| Percent of redundant areas [% area redundancy] | [0.61, 0.90] | 0.803 | 0.099 |
| Total number of fixations [n fixations] | [487, 1254] | 836.875 | 248.288 |
| Fixations to Saccade ratio [fixation : saccade] | [5.03, 12.07] | 7.560 | 2.813 |

* TP stands for the Markov transition probability between modes.

***Table 8.*** *List of evaluation metrics derived from inspection results.*

| Metric Name (units) [Short name] | Range | Mean | Std |
|---|---|---|---|
| Time spent to find 20% of defects (s) [t20] | [71.98, 383.32] | 203.08 | 107.80 |
| Time spent to find 50% of defects (s) [t50] | [156.86, 614.40] | 425.03 | 181.64 |
| Discovery rate (Defects found / s) [Defects : s] | [0.01, 0.05] | 0.0229 | 0.0096 |
| Total inspection time (s) [t total] | [374.03, 922.97] | 802.04 | 182.90 |
| Total defects found with image [n found] | [13, 18] | 16.875 | 1.727 |
| Total defects seen with gaze [n seen] | [17, 23] | 20.375 | 2.134 |
| Spatial ability score [PSVT:R] | [14, 29] | 24.750 | 4.979 |
| Percent area of bridge covered [$A_{covered}$] | [0.58, 0.85] | 0.743 | 0.093 |

### 3.4 Descriptor-performance correlation analysis

A descriptor-metric correlation analysis was used to evaluate whether the extracted behavioral descriptors were associated with inspection performance. Because the study included eight inspectors, the descriptor–performance analysis was treated as an exploratory and hypothesis-generating feasibility assessment rather than a confirmatory statistical test. The goal was to identify candidate behavioral descriptors that may be associated with inspection performance and warrant evaluation in larger studies. To reduce overinterpretation, the analysis reported 80% confidence interval (CI), leave-one-user-out mean absolute error for linear prediction, and permutation-based two-tailed Spearman correlation tests. The permutation test provided a nonparametric assessment of association, while the leave-one-user-out error quantified how well each descriptor generalized when predicting the performance of an unseen inspector.

Table 9 summarizes the descriptor-metric pairs that showed stronger exploratory associations based on the Spearman's coefficient, $\rho$, and two-tailed hypothesis test p-value, defined as $\rho^2 > 0.6^2\ and\ p < 0.1$. A total of 21 stronger trends were identified. To provide a more complete record of the screening analysis, weaker exploratory associations, defines as $\rho^2 > 0.4^2\ and\ p < 0.25$ , are reported in Appendix Table A1. Overall, the observed associations suggest that inspection-mode descriptors may contain information related to inspection performance; however, the small sample size and limited variability in performance metrics prevent definitive conclusions about effect size or generalizability.

***Table 9.*** *List of descriptor-metric pairs with stronger correlations.*

| Descriptor | Metric (units) | ρ | $p$ | 80% CI | LOUO MAE |
|---|---|---|---|---|---|
| μ loc dwell | PSVT:R | -0.699 | 0.065 | [-0.893, -0.284] | 3.883 |
| μ glob dwell | PSVT:R | -0.855 | 0.01 | [-0.952, -0.606] | 2.949 |
| t nav | PSVT:R | 0.723 | 0.053 | [0.328, 0.903] | 3.334 |
| t loc | t50 (s) | 0.714 | 0.058 | [0.312, 0.899] | 121.038 |
| ↳ | t total (s) | 0.643 | 0.099 | [0.188, 0.871] | 140.106 |
| t glob | n seen | -0.831 | 0.015 | [-0.943, -0.551] | 1.071 |
| ↳ | n found | -0.856 | 0.011 | [-0.952, -0.607] | 1.346 |
| nav → loc | t20 (s) | 0.905 | 0.004 | [0.728, 0.969] | 64.63 |
| ↳ | t50 (s) | 0.714 | 0.056 | [0.312, 0.899] | 182.043 |
| ↳ | $A_{covered}$ | 0.695 | 0.057 | [0.276, 0.892] | 0.119 |
| n fixations | t total (s) | 0.786 | 0.032 | [0.452, 0.927] | 142.948 |
| ↳ | Defects : s | -0.881 | 0.009 | [-0.961, -0.668] | 0.007 |
| loc : glob | n found | 0.677 | 0.078 | [0.245, 0.885] | 1.437 |
| $H_m$ | PSVT:R | 0.651 | 0.089 | [0.201, 0.874] | 4.437 |
| glob → nav | PSVT:R | 0.807 | 0.022 | [0.498, 0.934] | 3.416 |
| glob → loc | t20 (s) | 0.762 | 0.039 | [0.403, 0.918] | 61.671 |
| ↳ | t50 (s) | 0.714 | 0.063 | [0.312, 0.899] | 165.161 |
| fixation : saccade | n found | 0.894 | 0.006 | [0.700, 0.965] | 1.278 |
| ↳ | n seen | 0.735 | 0.047 | [0.351, 0.907] | 1.921 |
| % area redundancy | PSVT:R | 0.663 | 0.078 | [0.221, 0.879] | 2.753 |

Several associations in Table 9 are behaviorally interpretable and support the value of the proposed inspection mode time series. The most important finding is that transition-based descriptors were associated with inspection efficiency. In particular, the transition probability from navigation to local inspection was positively associated with $t_{20}$, meaning that inspectors who more frequently moved directly from navigation to focused assessment tended to find early defects more slowly. This may indicate that a scan-before-inspect pattern is more efficient than moving directly from navigation into local evaluation, because broader scanning may help inspectors prioritize candidate defect locations before committing to detailed assessment. Time spent in global scanning was also negatively associated with both the number of defects seen and the number documented. This suggests that although global scanning is necessary for orientation and search, excessive time in this mode may reduce opportunities for focused defect assessment. This finding supports the distinction between global scanning and local inspection: both involve visual attention, but they appear to have different relationships with inspection completeness.

Conventional fixation-based descriptors showed useful but less specific relationships. The fixation-to-saccade ratio was positively associated with both defects seen and defects documented, suggesting that stable visual processing supports defect recognition. However, total fixation count was also associated with total inspection time, which illustrates a limitation of cumulative gaze metrics: they can confound inspection strategy with task duration. This reinforces the benefit of the inspection mode time series, which captures the temporal organization of behavior rather than only the total amount of gaze activity.

Finally, several inspection-mode descriptors were associated with the PSVT:R spatial ability. Higher PSVT:R scores were linked to shorter global-scanning dwell times and higher transition probability from global scanning to navigation. This may suggest that inspectors with stronger spatial ability orient more quickly and reposition more readily, although this interpretation should be treated cautiously because spatial ability was not clearly associated with inspection efficiency in this study.

Overall, these exploratory associations suggest that dwell time, transition probability, and other descriptors derived from the inspection mode time series capture meaningful aspects of inspection strategy. However, given the small sample size and limited variability in performance, these findings should be interpreted as hypotheses for future validation rather than definitive evidence of predictive relationships.

## 3.5. Discussion, limitations, and implications

The results of this study demonstrate the feasibility of inspection mode temporal windowing to analyze visual attention and behavior in unconstrained 3D bridge inspection. The VISTA platform supported synchronized data collection and qualitative visualization of inspector behavior, such as replay analysis, gaze heatmaps, and area revisit heatmaps. These tools helped characterize redundant inspection behavior, identify defects observed but not documented, and compare differences in spatial attention. The inspection mode time series further enabled behavioral segmentation of inspection sessions and supported the extraction of higher-level descriptors such as dwell-time, transition probabilities, and transition entropy.

The main contribution of the proposed approach is that it converts raw gaze and movement data into a behaviorally interpretable signal. This signal can be analyzed using standard time series signal processing techniques, such as Markov transition analysis, allowing inspection behavior to be compared across users without relying on manually defined AOIs or constrained inspection paths. The feature analysis and model performance results show that global scanning, local inspection, and navigation have distinct and interpretable signatures. The correlation analysis further suggests that descriptors derived from these modes may be associated with inspection speed, completeness, spatial ability, and redundancy.

However, several limitations affect the interpretation of these results. The first limitation is the small sample size. With eight inspectors, the correlation analysis cannot provide conclusive evidence of relationships between behavioral descriptors and inspection performance. The reported associations should therefore be interpreted as hypotheses for larger studies rather than definitive generalizable findings.

The second limitation concerns the inspection task itself. Most users identified a similar subset of defects, and many defects appeared to be visually salient once they entered the participants' view. This is consistent with the Guided Search model [4], in which, when the target differs strongly from the background and distractors are visually similar, bottom-up guidance can make targets easier to find. In the present bridge model, the concrete texture was relatively uniform, while many cracks and surface defects had a strong contrast against the background. As a result, participants may have relied heavily on bottom-up salience rather than more realistic top-down inspection strategies based on structural hierarchy, deterioration mechanisms, or defect-prone regions. This reduces ecological validity and limits variability in defect-detection performance.

The hairline cracks provide an important exception. These defects were not documented by any participant and required closer viewing to detect. Their difficulty suggests that future studies should include more subtle defects, more realistic distractors, variable lighting, surface discoloration, occlusions, and defects placed according to realistic deterioration patterns. Such improvements would make the task more representative of real bridge inspection and would likely create greater variability in inspection strategy and performance.

The third limitation is related to fixation and attention modeling in 3D environments. The I-VT fixation filter and fixed-size spotlight model provide practical and reproducible approximations, but they simplify the complexity of visual attention in navigable 3D scenes. Velocity-based fixation filters may group gaze samples that are close in visual angle but correspond to different depth planes. Similarly, the fixed-size spotlight model does not account for changes in attentional extent across users, tasks, or levels of cognitive demand. Future work should investigate fixation classification and attention-field estimation methods designed specifically for dynamic 3D environments.

The fourth limitation concerns supervised inspection-mode labeling. The inspection modes used in this study were defined through theory and refined through consensus discussion among labelers. This process produced usable labels for training, but it also introduced a degree of subjectivity. Preliminary labeling trials showed low agreement before discussion, suggesting that different labelers may interpret the boundaries between global scanning, local inspection, and navigation differently. Consensus labeling reduced this inconsistency but does not eliminate the possibility that the learned modes reflect the labeling group's operational definitions. Future work should investigate larger labeling groups, formal reliability testing, semi-supervised learning, and unsupervised or self-supervised representation learning. Such methods may help identify behavioral states directly from data while allowing humans to interpret the resulting patterns.

Despite these limitations, the study demonstrates that automated gaze-based behavioral segmentation is a promising approach for evaluating bridge inspection behavior in unconstrained 3D environments. In inspector training applications, the method could be used to provide feedback to novice inspectors by identifying excessive navigation, insufficient local assessment, limited spatial coverage, or redundant revisiting of the same regions. In infrastructure maintenance applications, the resulting descriptors could support more objective documentation of inspection process quality, complementing defect-level outcomes with measures of how the inspection was performed. In research applications, VISTA provides a controlled environment for studying human factors in Unmanned Aerial System (UAS)-assisted inspection and for developing human-aware inspection protocols. The resulting behavioral descriptors may also support future work on automated UAS inspection systems by providing human-derived benchmarks for efficient and thorough inspection behavior.

## 4. Conclusion

This paper presents an automated framework for converting unconstrained 3D bridge inspection behavior into an inspection mode time series without relying on predefined areas of interest or constrained navigation paths. The proposed approach represents an inspection as a temporal sequence of functional behavioral states, including global scanning, local inspection, and navigation. This representation provides a structured and interpretable way to analyze how inspectors move between broad visual search, focused defect assessment, and movement control during a realistic inspection task. By transforming raw gaze, head movement, drone navigation, and scene-geometry data into a behavioral time series, the framework enables inspection behavior to be analyzed using higher-level behavioral descriptors, such as dwell times, transition probabilities, transition entropy, and mode duration, that reveal meaningful differences in inspection strategies not observable through traditional fixation-based metrics alone. Unlike conventional gaze-analysis methods that rely on manually defined areas of interest, constrained viewing conditions, or aggregate fixation statistics, the proposed temporal-windowing framework can operate in unconstrained 3D environments where inspectors freely navigate, change viewpoint, and examine complex structural components. This makes the approach suitable for studying inspection behavior in settings that better reflect emerging remote, UAS-assisted, and virtual inspection workflows.

The feasibility of the proposed framework was demonstrated using the VISTA platform, which supported synchronized collection of gaze, camera, movement, and inspection-performance data. Supervised classifiers were trained to identify inspection modes from short temporal windows. The resulting inspection mode time series captured interpretable behavioral patterns across participants, including differences in time spent navigating, scanning, and locally inspecting, as well as differences in transition structure and behavioral persistence. The results showed strong classification performance under random cross-validation and moderate generalization under leave-one-user-out validation, indicating that the proposed inspection modes have distinguishable feature signatures while also reflecting user-specific variability.

The inspection mode time series also enabled the extraction of behavioral descriptors associated with inspection performance in an exploratory analysis. Transition-based descriptors, dwell-time measures, fixation-based descriptors, and redundancy measures showed interpretable relationships with metrics such as early defect discovery, number of defects found, spatial coverage, and spatial ability. These findings suggest that the temporal organization of inspection behavior contains information that is not captured by conventional aggregate gaze metrics alone. In particular, descriptors based on transitions among navigation, global scanning, and local inspection provide a way to characterize inspection strategy, rather than only measuring the amount of visual activity. In addition, this study introduced an area-revisit heatmap to quantify repeated spatial coverage during inspection. This descriptor complements surface-based gaze heatmaps by capturing when inspectors return to previously viewed regions after a period of absence. The area-revisit analysis provides insight into inspection redundancy, repeated verification, and potentially inefficient re-examination of the same regions. Together with the inspection mode time series, this spatial-temporal representation offers a richer basis for evaluating inspection behavior and comparing inspection strategies across users.

The findings should be interpreted in light of several limitations. The user study included a small number of participants, so the descriptor-performance correlations should be viewed as exploratory hypotheses rather than confirmatory evidence. The inspection environment also included defects that were visually salient once they entered the inspector's field of view, which may have reduced the need for expert top-down search strategies and limited variability in performance. In addition, the fixation filter, fixed spotlight model, and supervised labeling process provide practical approximations but do not fully capture the complexity of attention allocation in dynamic 3D environments.

Future work should validate the proposed framework with a larger and more diverse participant pool, including certified bridge inspectors and novice trainees. Future inspection environments should also include more realistic defect distributions, subtle deterioration patterns, visual distractors, occlusions, and variable lighting conditions. Further methodological development is needed to improve fixation classification, attention-field estimation, and behavioral-state discovery in 3D environments. Semi-supervised, unsupervised, or self-supervised learning approaches may reduce the dependence on manually labeled inspection modes and may reveal additional behavioral states beyond those defined in this study.

Overall, this study demonstrates that converting inspection behavior into an inspection mode time series provides a promising foundation for objective, scalable, and interpretable analysis of bridge inspection processes. The proposed framework advances gaze-based inspection research from static or aggregate attention measures toward temporal behavioral modeling. More broadly, the work contributes to the automation of infrastructure inspection and maintenance by integrating computer-aided simulation, eye-tracking data, machine-learning classification, and process-level performance descriptors. This capability can support inspector training, human-performance assessment, human-UAS interaction research, and the development of human-informed automated inspection systems for civil infrastructure management.

## Data Availability

This study was approved by the Institutional Review Board (IRB) at the University of Illinois Urbana-Champaign (IRB protocol 24-1353). Due to our commitment to data anonymization, raw data will not be made available; however, statistical measures, visualizations, and trained models will be provided upon request.

## Acknowledgements

The authors acknowledge graduate students Yuxiang Zhao, Ishfaq Aziz, and Mohammadreza Rajaee for their assistance throughout the study and thank the anonymous volunteers who participated in the user study.

## Appendix

### Scoring of defects found during inspection

In order to determine the time and number of defects seen with gaze and found with image a scoring person, or rater, manually reviewed each user's inspection data, making use of different visualizations and their own judgment. To determine the time that a defect is found the rater looked at the images taken by the inspector, if a defect is close in distance and clearly framed within the center third of the image (using a nine-grid layout), then the defect was recorded as being found at the time that the image was taken, this also implies that the defect was seen with gaze. This rating scheme ensures consistency and reduces bias in scoring; however, some edge cases may exist, potentially compromising reliability. One example is if the inspector takes an image that accidentally captures a defect that they had not seen. On the other hand, an inspector may have found a defect and take an image that they believe to have sufficient quality but is in fact too far from the defect or does not properly frame the defect. None of these cases were perceived by the rater for this study, but it should be noted that alternative scoring methods could have been used to improve reliability.

To determine if a defect was not seen at all or if it was seen with gaze but not found or reported with an image the rater carefully reviewed the qualitative visualization tools such as the inspector view replay, the gaze point cloud, and the gaze heatmaps to make this determination for each of the defects that were not found with an image. Figure A1 shows an example of the latter scenario. In this case, the defect was counted as observed but not identified via image capture. The replay view and 3D gaze-point visualization provide complementary evidence: the replay shows the temporal context of the participant's attention, showing that the inspector's camera clearly framed the defect at some point during the inspection, and their gaze remained in the area for some time. Meanwhile, the gaze-point cloud shows whether gaze was concentrated around the defect location, showing that although there is a small concentration of gaze in the defect area it is not significant. These two visualizations signal that the defect was seen during scanning of the area, but an image was not captured likely because the defect went undetected.

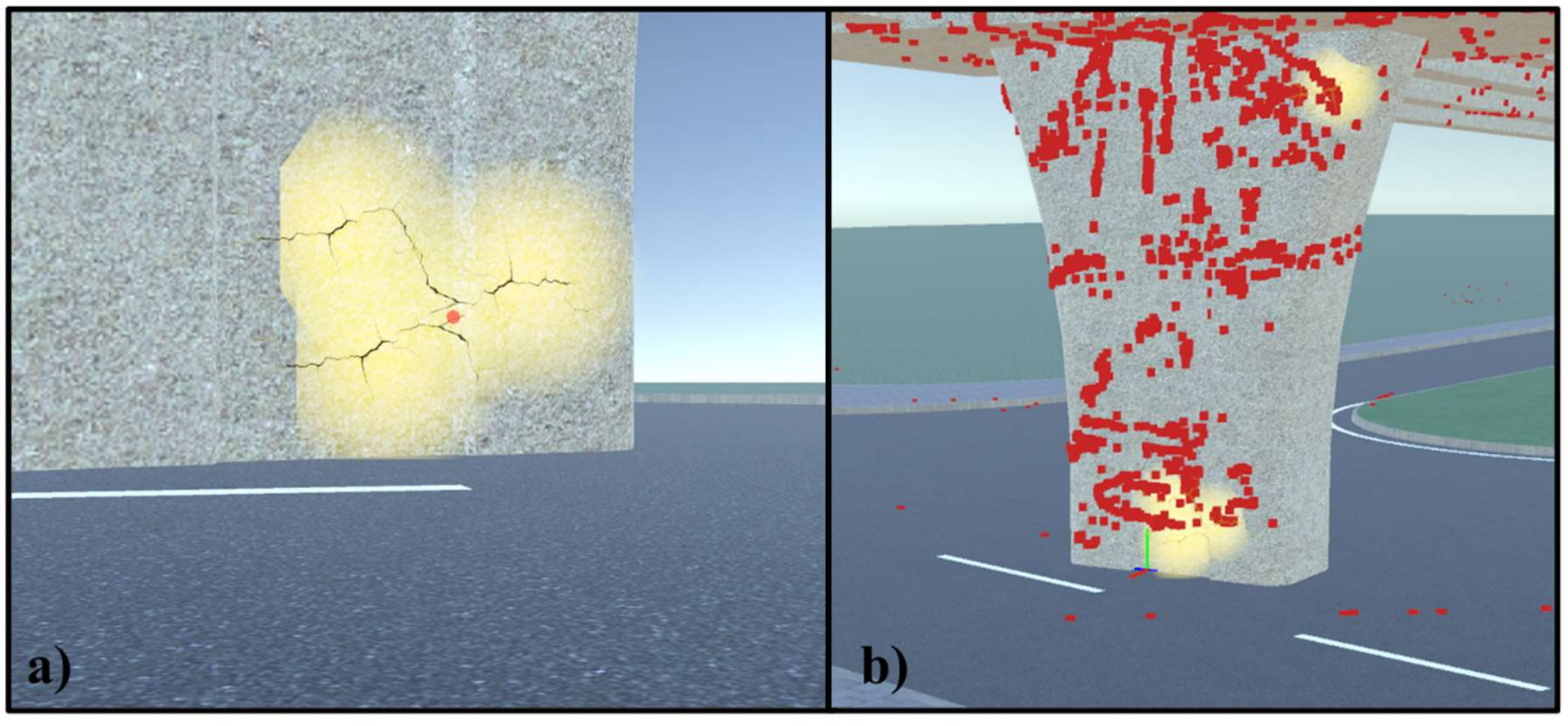

***Fig. A1.*** *Example of user 5 finding a defect but failing to document it. (a) Replay of inspector's point of view shows that their gaze, shown in red, was fixated at the crack location. (b) Gaze point cloud shows that the column was thoroughly inspected but no image was taken.*

**The effect of revisit-time threshold on area revisit heatmap**

Figure A2 evaluates the effect of the revisit-time threshold on the area revisit heatmap. A threshold of 20 seconds was selected as a compromise between two limiting cases. When the threshold approaches zero, repeated viewing is counted almost continuously, and the revisit heatmap becomes similar to the gaze heatmap. When the threshold is very large, few regions qualify as revisits, and the map approaches a first-visit representation. The 20 seconds threshold reduced the influence of continuous local viewing while preserving meaningful returns to previously inspected regions. This choice was also supported by the inspection-mode time series: inspectors remained in local inspection for more than 20 consecutive seconds only 7.6% of the time, and remained in any single inspection mode for 20 consecutive seconds only 4.2% of the time.

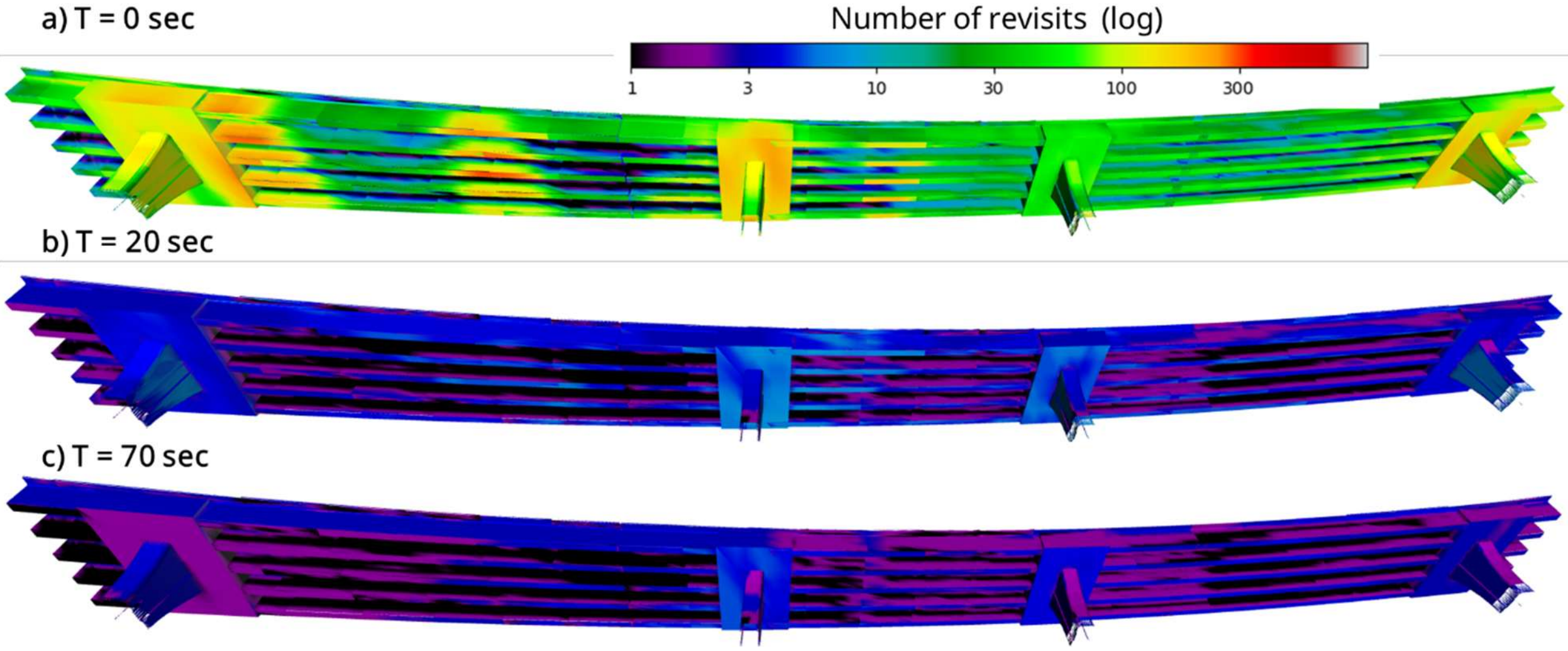

***Fig. A2.*** *Area revisit heatmap for user 3 with time thresholds (a) 0s, (b) 20s, (c) 70s*

**List of weekly correlated descriptor-metric pairs**

A weaker set of correlation results is presented in Table A1. These results correspond to relationships that may emerge more strongly in larger studies and could support some of the discussion topics in this study.

***Table A1.*** *List of weakly correlated descriptor-metric pairs.*

| Descriptor | Metric (units) | ρ | *p* | 80 CI | LOUO MAE |
|---|---|---|---|---|---|
| % area redundancy | n seen | 0.506 | 0.208 | [-0.016, 0.811] | 1.4 |
| fixation : saccade | Defects : s | 0.571 | 0.15 | [0.076, 0.840] | 0.008 |
| ↳ | $A_{covered}$ | 0.515 | 0.197 | [-0.004, 0.815] | 0.063 |
| ↳ | t total (s) | -0.5 | 0.222 | [-0.808, 0.024] | 152.322 |
| glob → loc | t total (s) | 0.5 | 0.224 | [-0.024, 0.808] | 162.76 |
| glob → nav | t total (s) | -0.524 | 0.203 | [-0.819, -0.008] | 128.955 |
| ↳ | Defects : s | 0.5 | 0.214 | [-0.024, 0.808] | 0.008 |
| loc : glob | t50 | 0.595 | 0.126 | [0.112, 0.851] | 153.708 |
| ↳ | t20 | 0.548 | 0.164 | [0.042, 0.830] | 87.768 |
| loc → glob | n found | -0.6 | 0.14 | [-0.853, -0.120] | 3.08 |
| n fixations | t50 | 0.571 | 0.14 | [0.076, 0.840] | 160.046 |
| ↳ | n found | -0.549 | 0.163 | [-0.831, -0.044] | 1.313 |
| ↳ | PSVT:R | -0.53 | 0.178 | [-0.822, -0.017] | 4.1 |
| ↳ | t20 | 0.524 | 0.202 | [0.008, 0.819] | 103.758 |
| t glob | A covered | -0.491 | 0.227 | [-0.804, 0.036] | 0.099 |
| ↳ | PSVT:R | -0.47 | 0.242 | [-0.794, 0.063] | 4.197 |
| t loc | t20 | 0.571 | 0.144 | [0.076, 0.840] | 98.602 |
| ↳ | Defects : s | -0.476 | 0.244 | [-0.797, 0.055] | 0.008 |
| t nav | t total | -0.595 | 0.131 | [-0.851, -0.112] | 130.259 |
| ↳ | Defects : s | 0.571 | 0.151 | [0.076, 0.840] | 0.008 |
| ↳ | t50 | -0.548 | 0.166 | [-0.830, -0.042] | 131.6 |
| μ glob dwell | n seen | -0.53 | 0.191 | [-0.822, -0.017] | 1.329 |
| ↳ | Defects : s | -0.5 | 0.215 | [-0.808, 0.024] | 0.008 |
| μ nav dwell | t50 | -0.595 | 0.139 | [-0.851, -0.112] | 137.891 |
| ↳ | t total | -0.476 | 0.242 | [-0.797, 0.055] | 130.416 |